\documentclass[a4paper, 12pt, one column, aasmacros]{article}

\usepackage[english]{babel}
\usepackage[utf8x]{inputenc}
\usepackage[T1]{fontenc}
\usepackage{aasmacros}
\usepackage{color}

\usepackage[top=1.3cm, bottom=2.0cm, outer=2.5cm, inner=2.5cm, heightrounded,
marginparwidth=1.5cm, marginparsep=0.4cm, margin=2.5cm]{geometry}

\usepackage{graphicx} 
\usepackage[colorlinks=False]{hyperref} 
\usepackage{amsmath}  
\usepackage{amsfonts} %
\usepackage{amssymb}  %
\usepackage{amsthm}
\usepackage{siunitx}

\DeclareMathOperator*{\argmin}{arg\,min}
\DeclareMathOperator{\e}{\mathrm{e}}

\newcommand{\figsizeeps}{width=\dimexpr 0.75\textwidth\relax}
\newcommand{\figsizepdf}{width=\dimexpr 0.84\textwidth\relax}

\title{Robust One-Step Estimation of Impulsive Time Series}
\author{H{\aa}kan Runvik, Alexander Medvedev}
\date{}
\begin{document}

\maketitle

\begin{abstract}
The paper deals with the estimation of a signal
model in the form of the output of a continuous linear time-invariant system
driven by a sequence of instantaneous impulses, i.e. an impulsive time series.
This modeling concept arises in, e.g., endocrinology when episodic hormone secretion events and elimination rates  are simultaneously estimated from sampled hormone concentration measurements. The pulsatile secretion is modeled with a train of Dirac impulses constituting the input to a linear plant, which represents stimulated hormone secretion and elimination. A previously developed one-step estimation algorithm effectively resolves the trade-off between data fit and impulsive input sparsity. The present work improves the algorithm so that it requires less manual tuning and produces more accurate results through the use of an information criterion. It is also extended to handle outliers and unknown basal levels that  are commonly recognized issues in biomedical data. The algorithm performance is evaluated both theoretically and experimentally on synthetic and clinical data.
\end{abstract}

\section{Introduction}
Estimating the parameters of a dynamical model from measured data is a fundamental part of system identification. Typically, knowledge of both input and output signals is assumed, but in some applications, the input can neither be controlled, nor measured. In such cases, the identification is a matter of time-series estimation. The present paper considers a particular setup of this type, where the input signal consists of a sequence of impulses, and the parameters and the input signal are estimated from sampled measurements of the model output, which consequently are termed as an impulsive time series. 

Impulsive time series estimation is particularly relevant in biological contexts, where systems driven by intrinsic impulsive feedback mechanisms that are hard to measure or model sometimes occur. Typical examples include gait models \cite{ppk01}, muscle activation \cite{lby02}, and population models in ecology \cite{ll17}. We will focus on a problem in endocrinology, where the secretion events and clearance rates of hormones are estimated from blood concentration measurements. In the traditional approach for solving this problem, each secretion event is represented by a function of a predefined shape, while a linear hormone clearance model is used. As a result, the output of the system is given by a convolution integral, and the model estimation constitutes a deconvolution problem. Software utilizing deconvolution include AutoDecon \cite{jpv09} and WINSTODEC \cite{spc02}; see also the overview in \cite{dl93}. Other methods that have been proposed include the Bayesian approach in \cite{j03} and the constrained least squares formulation in \cite{MM:13}. The present method builds upon the latter work where   Dirac impulses, rather than continuous functions, are utilized to represent the pulsatile secretion.

Due to physiological, ethical, and experimental limitations, the sampling rate of clinical endocrine data is often low compared to the half-life times of the involved substances. Combined with uncertainties in both measurements and models, this leads to a challenging estimation problem. There is in particular a fundamental trade-off between impulsive input sparsity and fit to data that needs to be addressed, regardless of which estimation technique that is employed. For example, the deconvolution methods mentioned above implement statistical tests (AutoDecon) and regularization (WINSTODEC) to avoid overfitting.

The goal of the current work is to formulate and solve the combined parameter and input estimation problem in a way that, on the one hand, imposes minimal additional assumptions, heuristics or manual tuning, and on the other hand, is feasible when faced with the challenges related to clinical data, such as measurement outliers and unknown basal levels. A hybrid model, i.e., a model where both continuous and discrete dynamics are included, turns out to be beneficial for this goal. The model we use is based on the closed-loop model of testosterone regulation introduced in \cite{MCS06,cms09}, with the feedback mechanism disregarded in the current work. The secretion events are represented by instantaneous impulses in this model. Naturally, such impulses are mathematical constructs that do not occur in real biological systems, but when the duration of the secretion bursts is significantly shorter than the sampling time of the series,  this approximation is motivated. Here it leads to a tractable mathematical formulation and enables the use of a one-step estimation method that was introduced in \cite{rm22}.

In statistics, one-step estimation methods refer to estimators where a preliminary estimate is improved upon by performing a single step of Newton's method, rather than the more common situation where this is recursively repeated until convergence. The motivation is that, under certain conditions, the asymptotic properties of the estimator do not improve by taking multiple steps (see, e.g., \cite[Ch.~5.7]{Vaart98}). Our method is also based on performing a single Newton step, but the motivation is different. Here it is employed to address the ill-posedness of the problem, by finding an initial estimate of the elimination rates such that the estimated impulsive input is guaranteed to be sparse, and then using the Newton step to refine this estimate while preserving the input sparsity.

An earlier version of the method considered here was used in \cite{rm22b} to estimate the elimination rates and secretion events in luteinizing hormone (LH) data sets collected from healthy males.
That work showed the promise of the method, but also revealed a number of limitations when it was applied to clinical data. The main contribution of this paper is to address the shortcomings of the algorithm, by adding the following features to the estimation method:
\begin{itemize}
    \item Point estimates for all parameters, obtained by extending the estimation algorithm with a novel regularization method and an information criterion;
    \item Basal level estimation through a direct generalization of the one-step algorithm;
    \item Robustness against  measurement outliers in the data through the incorporation of a robust least squares solver;
    \item Detection of outlying hormone profiles.
\end{itemize}

The rest of the paper is outlined as follows. In Section~\ref{sec:estmodel}, the model and estimation problem are introduced and the general estimation strategy and its application to a first-order system are provided. In Section~\ref{sec:endest}, the implementations of the above listed features are presented. The resulting estimation algorithm is applied to synthetic and clinical data in Section~\ref{sec:exp}, followed up by discussion and conclusions in Section~\ref{sec:concl}.

\section{Model and estimation problem}\label{sec:estmodel}
The impulsive time series is defined as the (possibly irregularly) sampled output measurements $y(t_k), k=1,\dots,K$ of a system of the form
\begin{equation}\label{eq:sysIV}
    y(t) = y_0 + G(p) \xi(t),
\end{equation}
where $G(p)$ is a linear time-invariant single-input single-output system with $p$ denoting the differential operator, $y_0$ is a constant offset, and $\xi(t)$ a sequence of time-shifted Dirac delta-impulses with positive weight, i.e.
\begin{equation*}
    \xi(t)=\sum_{n=0}^{\infty}d_n \delta\left(t-\tau_n\right),
\end{equation*}
where $d_n>0$, $\tau_n>0$. In endocrine applications, $y(t_k)$ is a hormone concentration measured in blood samples, $y_0$ is the basal level of this hormone, $\xi(t)$ represents the pulsatile hormone secretion, and $G(p)$ describes the linear elimination and stimulated secretion of the involved hormones. The operator $G(p)$ will thus admit a minimal state-space realization in the form of a compartmental model, and $\xi(t)$ will generally not be available for measurement.
The impulsive time-series estimation problem now consists of evaluating the weights $d_n$, the times $\tau_n$, $y_0$ and the parameters of $G$ from $y(t_k)$.

\subsection{Estimation strategy}\label{sec:eststrat}
The basis of the estimation strategy is a least squares formulation derived in \cite{MM:13}, which assumes a second-order $G(p)$ with a particular parametrization, which we describe in Section~\ref{sec:endest}. In this formulation, the linear plant parameters and the basal level form a vector $\omega$ which belongs to a set $\mathcal D_\omega \subseteq \mathbb R^{m}$ and is estimated by a least squares method, while the impulse times are assumed to coincide with the sampling times.

\begin{figure}
    \centering
    \expandafter\includegraphics\expandafter[\figsizeeps]{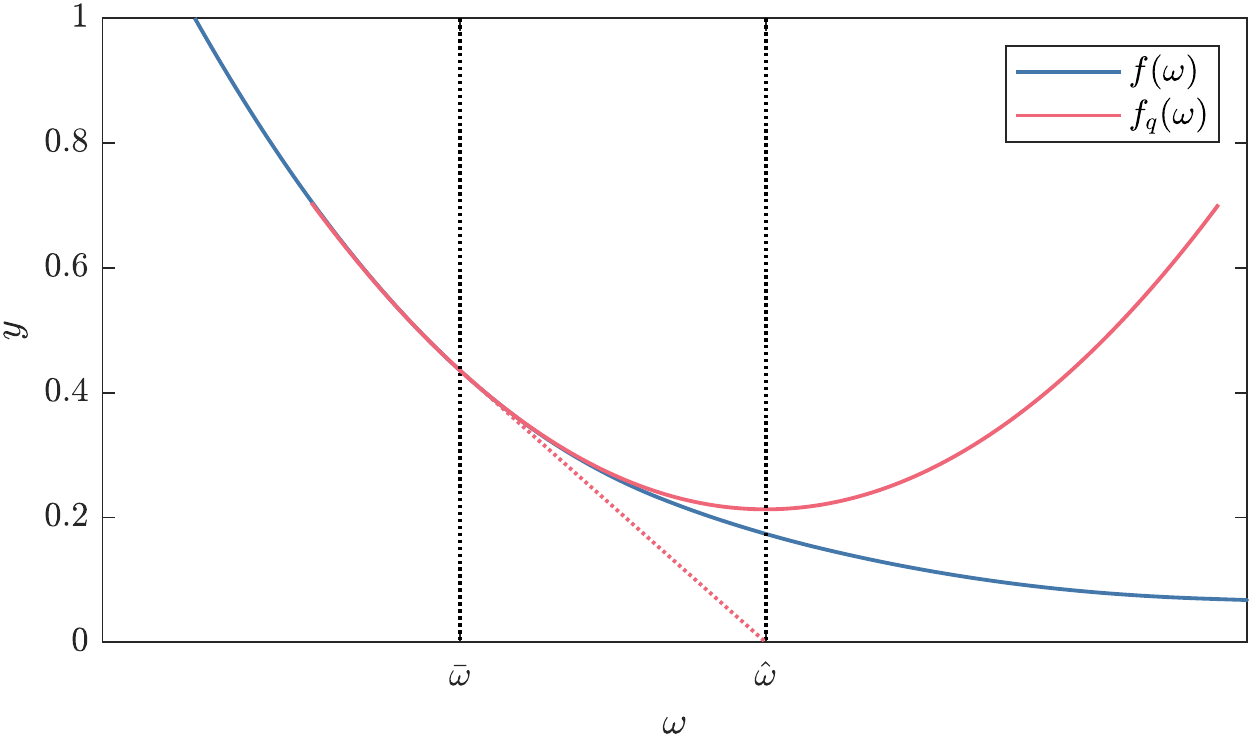}
    \caption{Residual sum of squares $f(\omega)$ decreases monotonously (blue line). Quadratic approximation at $\bar \omega$ gives a function $f_\mathrm q(\omega)$ (red line), whose minimum $\hat \omega$ is used as a parameter estimate. The tangent of $f(\omega)$ at $\bar \omega$ crosses zero at $\hat \omega$.}
    \label{fig:newtonex}
\end{figure}

The impulse weights and initial states of the system can then be collected in a vector $\theta$, which is estimated by
\begin{equation}
\begin{aligned}\label{eq:optIV}
    \hat \theta(\omega) = \argmin_\theta ||Y(\omega) - \Phi(\omega) \theta ||^2, \\
    \text{s.t. } \theta \ge 0,
    \end{aligned}
\end{equation}
where $Y(\omega)$ is given by 
\begin{equation*}
    Y(\omega) = \begin{bmatrix} y(t_1)-y_0 & \dots & y(t_K)-y_0 \end{bmatrix}^\intercal,
\end{equation*}
and the regressor $\Phi(\omega)$ is derived from the linear dynamics. We now concentrate out $\theta$ and consider the residual sum of squares
\begin{equation*}
    f(\omega)=\|Y(\omega) - \Phi(\omega) \hat\theta(\omega) \|^2.
\end{equation*}
It would seem like the estimation problem now could be solved by minimizing $f$ with respect to $\omega$. This is however not possible, as a perfect fit to any data can be obtained with a dense input and sufficiently fast linear dynamics, or a sufficiently low basal level. 
In \cite{rm22}, the following strategy was introduced to solve this problem. Consider the function
\begin{equation*}
     f^\dagger(\omega)=\|Y(\omega) - \Phi^\dagger(\omega) \hat\theta^\dagger(\omega) \|^2,
\end{equation*}
where $\Phi^\dagger(\omega)$ and $\hat\theta^\dagger(\omega)$ are restricted to only include elements corresponding to the true impulse times. This implies that minimizing $f^\dagger(\omega)$ will give a least squares solution $\hat \omega$, but, since the impulse times are unknown, so is $f^\dagger$.
However, if $f^\dagger(\omega)$ and $f(\omega)$ are approximately equal in a subset $S$ of $\mathcal D_\omega$, the estimation can be performed through the following steps:
\begin{enumerate}
    \item Find a suitable point $\bar \omega \in S$;
    \item Approximate $f(\omega)$ by its second-order Taylor expansion around $\bar \omega$, denoted $f_\mathrm q(\omega)$;
    \item Let $\hat \omega = \argmin_{\omega \in \mathcal D_\omega} f_\mathrm q(\omega)$;
    \item Determine the impulse times by \eqref{eq:optIV} with $\omega=\hat \omega$.
\end{enumerate}
Note that steps~$2$--$3$ correspond to a single step in Newton's method in optimization.
In the case of a scalar $\omega$, a candidate for $\bar \omega$ is obtained by
\begin{equation}\label{eq:omegabar}
    \begin{aligned}
    \bar \omega = \argmin_{\omega \in \mathcal D_\omega} N_f(\omega),\\
    \textrm{s.t. } df(\omega)/dx<0,
    \end{aligned}    
\end{equation}
where
\begin{equation*}
    N_{f}(\omega)= -\frac{f(\omega)}{df(\omega)/d\omega},
\end{equation*}
and the constraint prevents infeasible solutions with perfect fit to the data, which corresponds to $f(\omega)=df(\omega)/d\omega=0$.  
The motivation for using the point $\bar \omega$ is two-fold. It is firstly at a reasonable distance to $\hat \omega$, a property that will be elaborated upon in Section~\ref{sec:initest}. Secondly, it also simplifies the minimization of $f_\mathrm q$, as it removes the need of second derivatives in the calculation since
\begin{equation*}
    \frac{dN_f(\omega)}{d\omega}\bigg|_{\omega=\bar \omega} = 0
\end{equation*}
implies
\begin{equation*}
    \frac{df(\omega)/dx}{d^2f(\omega)/dx^2}\bigg|_{\omega=\bar \omega} = \frac{f(\omega)}{df(\omega)/dx}\bigg|_{\omega=\bar \omega} = N_f(\bar \omega),
\end{equation*}
which leads to the estimate
\begin{equation}\label{eq:omegahat}
    \hat \omega = \bar \omega + N_f(\bar \omega).
\end{equation}
As a consequence, the step in Newton's method in optimization coincides with a step in Newton's root finding algorithm, see Fig.~\ref{fig:newtonex}.
The main downside with this strategy is that \eqref{eq:omegabar} does not generalize in a natural way to a vector-valued  $\omega$.

\subsubsection{Estimating the impulses}
Steps~$1$--$3$ described above follow the strategy in \cite{rm22b}. However, for step~$4$, a refinement is proposed in the present work. In the presence of noise, the solution to \eqref{eq:optIV} with $\omega=\hat \omega$ is generally not  sparse, so one has to decide which elements of $\hat \theta$ correspond to nonzero impulse weights. In \cite{rm22b}, this was decided through manually chosen parameters, while an $\ell_1$-regularization was used in \cite{MM:13} for a similar purpose. But, recalling that the input is
sparse when $\omega=\bar \omega$,
the same sparsity can be enforced when $\omega=\hat \omega$. However, even when $\omega=\bar \omega$, the determination of nonzero impulse amplitudes is not obvious, particularly since numerical solutions to \eqref{eq:optIV} typically have no elements exactly equal to zero. To avoid such ambiguities, we suggest to regularize the solution at $\omega=\hat \omega$ so that the residual sum approximately equals $f_\mathrm q(\hat \omega)$.

\subsection{The case of first-order dynamics}\label{sec:firstorder}
As a preliminary case, the estimation of an impulsive time series with first-order continuous dynamics is considered in this section. This simplified setup is adopted mainly to enable a tractable analytical analysis of the estimation method, but versions of the methods presented here are also applied in the second-order case. The model is written in state-space form as
\begin{equation}\label{eq:firstorder}
    \dot x = -bx + \xi(t), \quad y=x,
\end{equation}
where the parameter $b>0$ is estimated along with the input signal. As a convention, we let $\omega$ be the estimated parameter and $b$ denote the true parameter value. The initial state is not included in the estimation since it can be represented by an impulse at time zero. The impulse times and amplitudes are also not uniquely identifiable, since an impulse between two sampling times can be simultaneously shifted in time and re-scaled while leaving the output unaffected.
The impulse times are therefore restricted to occur at the sampling times, which leads the regressor and parameters of \eqref{eq:optIV} to become
\begin{equation*}
    \Phi(\omega) = \begin{bmatrix}
        \e^{-\omega (t_1-t_1)} & 0 & \dots & 0 \\
        \e^{-\omega (t_2-t_1)} & \e^{-\omega (t_2-t_2)} & \dots & 0 \\
        \vdots & \vdots & & \vdots \\
        \e^{-\omega (t_K-t_1)} & \e^{-\omega (t_K-t_2)} & \dots & \e^{-\omega (t_K-t_K)}   
    \end{bmatrix},
\end{equation*}
and
\begin{equation*}
    \theta = \begin{bmatrix}
        d_1 & d_2 & \dots & d_K
    \end{bmatrix}.
\end{equation*}

\subsubsection{The initial estimate}\label{sec:initest}
The initial estimate $\bar \omega$ should satisfy two criteria. First, it should result in a sparse estimated input signal. As shown for a second-order system in \cite{rm22}, this is achieved when the elimination rate is sufficiently slow. To explain this further, let $\hat d_n(\omega)$ denote the elements of $\hat \theta(\omega)$ and consider the integrals
\begin{equation*}
    Z = \int_0^\infty y(t)~dt = \frac{1}{b} \sum_n d_n, \quad \int_0^\infty \hat y(t)~dt = \frac{1}{\omega} \sum_n \hat d_n(\omega),
\end{equation*}
where $\hat y$ is the output corresponding to the estimated impulse weights $\hat d_n(\omega)$ and the number of impulses is assumed to be finite. Under the natural assumption that
\begin{equation*}
    \int_0^\infty \hat y(t)~dt \approx Z,
\end{equation*}
the sum $\sum_n \hat d_n(\omega)$ is approximately proportional to $\omega$. A low value of $\omega$ then acts as an $\ell_1$-constraint, which gives rise to sparse solutions.

The second criterion is that $\hat \omega - \bar \omega$ should be small, for the quadratic approximation of $f$ to be accurate. We thus want $\bar \omega$ to be as large as possible, while keeping the input sparse. The consequences of determining $\bar \omega$ by solving \eqref{eq:omegabar} are analyzed next; derivations of the results are given in Appendix~\ref{app:firstordersens}. Consider the response of a single impulse with the weight $d$ for system \eqref{eq:firstorder}, with the output measurements subject to i.i.d additive Gaussian noise with variance $\sigma^2$. Equation \eqref{eq:omegabar} then leads to the approximate relations
\begin{equation}\label{eq:minNf}
    N_f(\bar \omega) = \hat \omega-\bar\omega \approx \sqrt{\frac{c_0}{c_2}}, 
\end{equation}
where $c_0=(K-1)\sigma^2$ and
\begin{equation*}
    c_2 = \sum_{k=1}^K \bigg(\frac{\partial \alpha_0}{\partial \omega}\Big|_{\omega=b}-t_k d\bigg)^2\e^{-2bt_k}, \quad  \alpha_0= d~\frac{\sum\limits_{k=1}^K \e^{-(b+\omega)t_k}}{\sum\limits_{k=1}^K \e^{-2\omega t_k}}.
\end{equation*}
In the limit when the noise variance goes to zero, $\hat \omega$ approaches $b$ and the approximation becomes strict. If equidistant sampling with a period of $T$ now is assumed, the point
\begin{equation}\label{eq:omegatilde}
    \tilde \omega = b-\sqrt{\frac{c_0}{c_2}}\frac{\e^{bT}\sqrt{\pi/2}}{\sqrt{(K-1)(1-\e^{-2bT})}}
\end{equation}
gives an approximation of an upper limit for $\bar \omega$ such that multiple impulse estimates are unlikely to appear. If $\hat \omega$ is close to the true parameter value, the condition $\bar \omega<\tilde \omega$ corresponds to
\begin{equation*}
    K\gtrapprox\frac{\pi \e^{2bT}}{2(1-\e^{-2bT})} + 1.
\end{equation*}
In the experiments in Section~\ref{sec:firstorderexp}, $T=0.5$ and $b$ is of the order of $1$, which gives $K\gtrapprox 7.8$.
The calculation of $\tilde \omega$ is based on several approximations and it should therefore not be interpreted as an exact bound. However, \eqref{eq:omegatilde}  still indicates that $\bar \omega$ is likely to give a sparse input signal if the sampling frequency is high and $K$ is large. Also note the square root of $K-1$ in this equation, which limits the sensitivity of $\tilde \omega$ with respect to $K$. As a result, $\tilde \omega-\bar \omega$ grows relatively slowly with $K$, i.e. $\bar \omega$ is not unnecessarily far from this limit.

\subsubsection{Sensitivity analysis}
Sensitivity analysis of the estimation method is presented here. For detailed derivations, see Appendix~\ref{app:firstordersens}.

For simplicity, the impulse response of \eqref{eq:firstorder} is again considered, with sampled measurements subject to Gaussian i.i.d noise. The parameter $c_2$ introduced above then relates to the Fisher information $\mathcal I(b)$ for the parameter $b$ and, by the Cram\'er-Rao bound, the variance of its estimate as
\begin{equation*}
    \mathcal I(b) = \frac{c_2}{\sigma^{-2}} \le \mathrm{Var}(\hat \omega).
\end{equation*}
Combined with \eqref{eq:minNf}, that leads to the approximate relation
\begin{equation} \label{eq:varineq}
    \mathrm{Var}(\hat \omega)\gtrapprox \frac{\big(N_f(\bar\omega) \big)^2}{K-1},
\end{equation}
which becomes a proper inequality when the noise level approaches zero, and an equality when $K$ furthermore  tends to infinity, under the assumption that a single (correct) impulse is estimated (which the analysis in Section~\ref{sec:initest} indicates should happen when $K\rightarrow \infty$). This indicates that a short Newton step and a large $K$ corresponds to a good estimate.

Since inequality \eqref{eq:varineq} only provides a lower bound and does not take the particular estimation method given by \eqref{eq:omegabar}, \eqref{eq:omegahat} into account, the accuracy of the one-step estimation method needs to be investigated. To this end, the sensitivity of the estimate with respect to higher order terms of $f$ is determined. Consider the case
\begin{equation}\label{eq:cubicf}
    f(\omega)=f_{c_3}(\omega) = c_0 + c_2(\omega-b)^2 + c_3(\omega-b)^3,
\end{equation}
i.e. a third-order term is introduced to represent deviations from the quadratic assumption that the estimation strategy builds upon. The first-order effect of this is an error that scales with $c_3$ as
\begin{equation}\label{eq:cubicferror}
    E = \frac{3c_0}{2c_2^2} c_3 + O(c_3^2).
\end{equation}
Since $c_0$ scales with the noise and $c_2$ scales with the impulse weight squared, the factor $3c_0/2c_2^2$ is typically small. For example, with $d=0.5$, $T = 0.5$, $\sigma=0.01$, $K=10$, which roughly correspond to the parameter values in the experiments in Section~\ref{sec:firstorderexp}, $c_0=\num{9e-4}$, $c_2=0.0906$ and $3c_0/2c_2^2=0.165$.
Deviations from the quadratic assumption and the use of the one-step estimation strategy are therefore not expected to contribute significantly to the overall uncertainty of the estimate, in particular if the noise level is low. This is also in line with the numerical experiment in Section~\ref{sec:firstorderexp}.

\subsection{Implementation}

\subsubsection{Optimization formulation}
The strict constraint on the derivative in \eqref{eq:omegabar} cannot be implemented numerically. We therefore use the modified formulation
\begin{equation}
\begin{aligned}\label{eq:omegabar2}
    \bar \omega = \argmin_{\omega \in \mathcal D_\omega} N_f(\omega),\\
    \textrm{s.t. } df(z)/dz\le0, \forall z \le \omega,
\end{aligned}    
\end{equation}
which restricts the search space to the region where $f(\omega)$ is monotonously decreasing.
However, experiments with synthetic data have  shown that this region is not restrictive enough, in that it sometimes permits estimates with unreasonably large values of $\bar \omega$, compared to the true parameter value. A constraint on the total number of estimated impulses was used in \cite{rm22b} to prevent this, but that requires the specification of a somewhat arbitrary threshold for counting in an impulse, which is undesirable. We instead suggest to add a small positive constant $\varepsilon$ to $f$,
which acts as a regularization. If it is set to be of the same order of magnitude as the noise variance, which in the endocrine case can be approximated through the measurement uncertainty, it has a negligible effect on $N_f(\omega)$ in the range of values of $\omega$ that is relevant for the estimation.

\subsubsection{Optimization solution}
For reasons discussed in Appendix~\ref{app:firstordersens}, the function $N_f$ can display multiple local minima. For simplicity, we therefore suggest gridding to solve \eqref{eq:omegabar2} and to approximate the derivative of $f$ using finite differences over the same grid.

\section{Robust endocrine estimation}\label{sec:endest}
We now turn to the model which was studied in \cite{MM:13,rm22,rm22b} and also is the main focus of this paper. It has the form
\begin{equation}\label{eq:sys2}
\dot x = \begin{bmatrix}
    -b_1 & 0 \\
    g_1 & -b_2
    \end{bmatrix} x + \begin{bmatrix}
    1 \\ 0
    \end{bmatrix}\xi(t), \quad y = y_0 + \begin{bmatrix} 0 & 1 \end{bmatrix} x,
\end{equation}
i.e. $G(p)$ in \eqref{eq:sysIV} is specialized here to a second-order system. It could represent a number of hormone axes, but has mainly been applied 
to hormones from the male reproductive axis, which also is the application we consider here. The states of the system then correspond to the concentrations of gonadotropin releasing hormone (GnRH) and luteinizing hormone (LH) and $b_1$ and $b_2$ represent their respective elimination rates. The coefficent $g_1$ describes the secretion rate at which  LH is stimulated by GnRH, but, since it is not uniquely identifiable, $g_1=1$ is assumed without loss of generality.

Two version of the vector $\omega$ are considered:
\begin{align*}
    \omega = &\omega_1 = \begin{bmatrix}
        b_1 & b_2
    \end{bmatrix}, \\
    \omega = &\omega_2 = \begin{bmatrix}
        b_1 & b_2 & y_0
    \end{bmatrix},
\end{align*}
respectively corresponding to a known (i.e. zero) and unknown constant basal level. The set $\mathcal D_\omega$ is assumed to be a hypercube in the corresponding coordinate space, i.e.~the parameters are restricted to intervals $I_{b_1},I_{b_2},I_{y_0}$. The expressions for $\theta$ and $\Phi(\omega)$, see \eqref{eq:optIV}, are given by
\begin{equation*}
    \Phi(\omega)=\begin{bmatrix} \varphi(b_1, b_2, t_1) & \dots & \varphi(b_1, b_2, t_K) \end{bmatrix}^\intercal,
\end{equation*}
where
\begin{equation*}
    \varphi(b_1, b_2, t_i)=\begin{bmatrix}\e^{b_2(t_i-t_1)} & z(b_1, b_2, t_i-t_1) & \dots & z(b_1, b_2, t_i-t_{K-1}) \end{bmatrix}^\intercal,
\end{equation*}
\begin{equation*}
    \theta=\begin{bmatrix}x_2(t_1) & d_1 & \dots & d_{K-1} \end{bmatrix}^\intercal,
\end{equation*}
\begin{equation*}
    z(b_1, b_2, t) = \frac{\e^{-b_2 t}-\e^{-b_1 t}}{b_1-b_2}H(t),
\end{equation*}
and $H$ is the Heaviside step function.

A difference compared to the first-order case is that impulses that occur between sampling times can be uniquely represented by impulses at the sampling times \cite{MM:13}. Furthermore, with  $\omega$ as a vector, it is no longer possible to use \eqref{eq:omegabar} directly to find an initial estimate. In \cite{rm22}, two approaches were presented to resolve this, when the basal level is fixed. In the case of very low measurement noise, \eqref{eq:omegabar} was used anyway, but with $N_f(\omega)$ defined with a partial derivative with respect to the slower elimination rate, and the minimization being performed over both parameters. With a higher noise level, gridding over one parameter and optimizing over the other produced a set of possible solutions, in the form of a curve $\gamma$ in the set $\mathcal D_\omega$ of the parameter space. In \cite{rm22b}, this curve was shown to coincide with the posterior distribution from a Markov-chain Monte-Carlo estimator, i.e. it represents a direction of high variance in $\mathcal D_\omega$. However, for practical applicability of the method, point estimates are also required when the noise level is higher.

\subsection{Point estimates}\label{sec:pointest}
A method for identifying point estimates along the curve $\gamma$ is presented in this section. For simplicity, the basal level is assumed to be known, while the case of unknown constant basal level is covered in the next section. The curve $\gamma$ is defined by all points $(b_1,\hat b_2(b_1)) \in I_{b_1} \times I_{b_2}$, where $\hat b_2(b_1)$ is
determined by
\begin{equation}\label{eq:b2bar}
\begin{aligned}
    \bar b_2(b_1) = \argmin_{b_2 \in I_{b_2}} N_f(b_1,b_2),\\
    \textrm{s.t. } \partial f(b_1,z)/\partial z\le0, \forall z \le b_2,
\end{aligned}
\end{equation}
\begin{equation*}
    \hat b_2(b_1) = \bar b_2(b_1) + N_f(b_1,\bar b_2(b_1)),
\end{equation*}
where
\begin{equation*}
    N_{f}(b_1,b_2)= -\frac{f(b_1,b_2)}{\partial f(b_1,b_2)/\partial b_2}.
\end{equation*}

Motivated by the analysis of the first-order dynamics in Section~\ref{sec:firstorder}, it is expected that estimates along $\gamma$ display a beneficial trade-off between input sparsity and fit to the data. To explain the method used to compare these estimates, more details on the impulse estimation procedure outlined in Section~\ref{sec:eststrat} are needed.

\subsubsection{Impulse estimation}
The impulses are estimated through regularization, where some estimated weights in $\hat \theta$ are set to zero. However, since the corresponding impulses  are constrained to the sampling times, whereas in reality, impulses will almost surely occur between these, the estimated weights are not used directly. Instead, the regularization is performed based on the impulses obtained by merging consecutive impulses, according to Algorithm~1 in \cite{MM:13}.

Let $m_k$ denote the weight of the impulse obtained by merging impulse $k$ and $k+1$. To determine which pairs that should be used to form each merged impulse (impulse $k$ could be combined with either impulse $k-1$ or $k+1$), the following linear integer programming formulation, which finds the combination which minimizes the total sum of the impulse weights, is used:
\begin{align*} 
    \hat P =\argmin_P D^\intercal P, \\
    \text{s.t. }p_{1,k}+p_{2,k-1}+p_{2,k}=1 \text{ for } k=2,\dots K,\\
    p_{1,1} + p_{2,1}=1,\\
    p_{i,k} \in \{0,1\} \text{ for } k=1,\dots K, i=1,2,
\end{align*}
where
\begin{align*}
    D^\intercal = \begin{bmatrix}
        d_1& \dots& d_K& m_1& \dots m_{K-1}
    \end{bmatrix},\\
    P^\intercal = \begin{bmatrix}
        p_{1,1}& \dots& p_{1,K}& p_{2,1}& \dots& p_{2,K-1}
    \end{bmatrix},
\end{align*}
and $p_{2,k}$ indicates that impulses $k$ and $k+1$ are merged, while $p_{1,k}$ indicates that impulse $k$ is not merged. By its construction, the minimization will tend to merge most impulses.

When the linear programming formulation is applied to the impulse estimates obtained from \eqref{eq:optIV} with the parameters $b_1,\hat b_2(b_1)$, the nonzero elements of $\hat P$ define a set of impulses. We now define the function $\hat f_n$ as the residual sum of \eqref{eq:optIV}, when the $n$ largest of these impulses are included, and let $\hat c_0$ denote the quadratic approximation of $f$ evaluated at $\hat b_2(b_1)$, i.e.
\begin{equation}\label{eq:c0}
    \hat c_0(b_1)=f_\mathrm q(b_1,\hat b_2(b_1)) = \frac{1}{2}\big(N_f(b_1,\bar b_2(b_1))\big)^2\frac{\partial^2 f(b_1,b_2)}{\partial b_2}\bigg|_{b_2=\bar b_2(b_1)}.
\end{equation}
The number of impulses to include is then given by
\begin{equation}\label{eq:n0}
    n_0(b_1) = \argmin_{n\in \mathbb N} |\hat f_n(b_1,\hat b_2(b_1)) - \hat c_0(b_1)|,
\end{equation}
i.e. for each $b_1$, the estimated input is given by the $n_0(b_1)$ largest impulses defined by $\hat P$, which results in the residual sum of squares $\hat f_{n_0(b_1)}(b_1,\hat b_2(b_1))$.

\subsubsection{Comparing estimates}
We now have parameters corresponding to the two criteria needed to evaluate the estimates along $\gamma$; $n_0(b_1)$ characterizes the input sparseness and $\hat f_{n_0(b_1)}(b_1,\hat b_2(b_1))$ the fit to data. Define the set $N=\{n \in \mathbb N \mid \exists b_1 \text{ s.t. } n_0(b_1)=n\}$. To obtain point estimates along $\gamma$, a first step is to determine the set of Pareto-efficient estimates
\begin{equation*}
    \hat b_1^{(n)} = \argmin_{b_1:n_0(b_1)=n} \hat f_n(b_1,\hat b_2(b_1)), \quad \hat b_2^{(n)} = \hat b_2(\hat b_1^{(n)}),
\end{equation*}
for $n\in N$.
As a further refinement, an information criterion (see e.g. \cite{ss04}) can be used to identify a single estimate. We use the Bayesian information criterion (BIC) for this purpose. Observing that $n$ impulses correspond to $2(n+2)$ estimated parameters, and assuming Gaussian i.i.d. noise, the number of impulse estimates given by this criterion is
\begin{equation*}
    n^\mathrm{BIC} = \argmin_{n\in N;n\le n_\mathrm{max}}\big\{ K\log\big( \hat f_n\big(\hat b_1^{(n)},\hat b_2(\hat b_1^{(n)})\big)\big) + 2(n+2) \log K\big\},
\end{equation*}
where $n_\mathrm{max} \in \mathbb N$ is an upper bound of the number of parameters. 
Finally, the corresponding estimated parameters are defined by 
\begin{equation*}
    \hat b_1^{\mathrm{BIC}} = \hat b_1^{(n^\mathrm{BIC})},\quad \hat b_2^{\mathrm{BIC}}=\hat b_2^{(n^\mathrm{BIC})}.
\end{equation*}
Note that the BIC requires the sample size to be much larger than the number of parameters, which generally does not hold in the present application. However, the results with the method have proven satisfactory in numerical experiments, as demonstrated in Section~\ref{sec:exp}. 

\subsection{Basal level}\label{sec:basal}
The estimation of the basal level $y_0$ suffers from the same problem as the estimation of the elimination rates, in the sense that the fit will always increase when the basal level is lowered and, as a result, more nonzero impulses are estimated.
To resolve this, we again propose a one-step estimation algorithm. So, if $b_1$ is assumed to be known, minimizing $N_f$ and taking the Newton step for a range of different basal levels produces a curve $\gamma$ in the $b_2$-$y_0$ plane, from which a point estimate can be obtained according to the strategy presented above. The problem is that $b_1$ is not known. Furthermore, the experiments with the Markov-chain Monte-Carlo estimates in \cite{rm22b} show that multiple parameter values can give a similar fit to data even when the basal level is fixed, so adding one more parameter may just add another dimension to the space of plausible solutions.

However, if one elimination rate is significantly faster than the other, i.e. $b_1\gg b_2$, the situation becomes more promising. In such cases, $b_1$ will have a smaller effect on the output (other than a pure re-scaling) and therefore be hard to estimate. On the other hand, this also implies that estimates of $b_2$ and $y_0$ should be insensitive to the value $b_1$. In this situation, performing the estimation for a number of fixed values of $b_1$ as outlined above will therefore give similar estimates of the basal level. To choose between these, an information criterion can again be employed. This choice would also correspond to an estimate of $b_1$, however the uncertainty here is expected to be significant.

Fortunately, in many hormone axes, the elimination rate of the releasing hormone is significantly faster than the elimination rate of the other hormones of the axis, so the proposed method has practical relevance. For example, the elimination rate of GnRH is an order of magnitude faster than the elimination rate of LH (see \cite{kv98}). 

\subsection{Robustness against outlying measurements}\label{sec:robust}
Measurement errors are often large in clinical endocrine data and the presence of outlying measurements is particularly problematic. This situation can be described statistically as a fraction of the measurement errors being drawn from a corrupting distribution, as opposed to Gaussian i.i.d. errors. It is well known that the performance of least squares estimates such as \eqref{eq:optIV} can be severely degraded under such circumstances.

In biomedical applications, robust estimation techniques are the recommended way to counteract this sensitivity, as opposed to simpler strategies based on e.g. residual analysis \cite{fv12}. AutoDecon does use the latter method for robustness, but the scheme is more involved as it includes repeated estimation steps \cite{jpv09}. The linear least squares formulation in our estimator enables the simple strategy of replacing \eqref{eq:optIV} with a robust least squares solution, if outliers in the data set are suspected. We use the robust risk minimization algorithm presented in \cite{ozs20} for this purpose. The algorithm produces a robust solution by reducing the effective sample size, which results in a down-weighting of the outliers and only requires the user to specify an upper bound $\epsilon$ on the fraction of corrupted data points. The method thus also offers an automatic and non-subjective method for detecting measurement errors in hormone data, as opposed to more ad hoc methods such as, e.g., the methods compared in \cite{vjr19}.

The introduction of the robust least squares solution requires two small modifications of the one-step algorithm. First, to have consistent finite-difference approximations of derivatives, the weighting of the data points in the robust least squares solutions must be consistent. The function values $f(b_1,b_2-h)$ and $f(b_1,b_2+h)$, which are used for \eqref{eq:b2bar} and \eqref{eq:c0}, are therefore calculated with the weights of the robust least squares solution for $f(b_1,b_2)$. Second, for \eqref{eq:n0} to be consistent, the weights from $f(b_1,\bar b_2(b_1))$ are used in the calculation of $\hat f_n(b_1,\hat b_2(b_1))$.

\subsection{Outlying hormone profiles}\label{sec:outlyingind}
Hormone profiles that are inconsistent with the chosen model structure are sometimes encountered. In the case of LH measurements,
there are several congenital or acquired conditions that can affect the functioning of the male reproductive axis (see, e.g., \cite{ccg16}), and thus disrupt the expected pulsatile behaviour. As previously mentioned, due to the ill-posedness of the estimation problem, it is in principle possible to find a good fit to the data even in such a case. But, that is typically not a relevant solution, since many other such solutions also exist and there is no way to choose one over another. Our estimation method instead indicates such cases by $\bar b_2$ in \eqref{eq:b2bar} coinciding with the lower boundary of $I_{b_2}$, so that the estimated elimination rate becomes very slow. An intuitive explanation of this behavior is that the variations in the measurements are interpreted by the algorithm as high-variance noise, rather than the responses to distinct impulses, and the slow elimination rate corresponds to the moving average of the signal being approximately constant.

\subsection{Optimization solution}
The estimation of $\gamma$ according to \eqref{eq:b2bar} is again based on gridding. The derivative is calculated using finite differences over the grid points in the non-robust case, while the robust implementation requires additional function evaluations for the derivative. To decrease the computation time, gridding over the full range of $b_1$ is only performed for a sparse subset of the $b_2$-values, with local optimization performed in between. For further details, we refer to the code provided online.

\section{Experiments}\label{sec:exp}
The performance of the one-step estimation method is demonstrated on synthetic and clinical data. For the former, Matlab code is available at \url{https://github.com/HRunvik/Robust-One-Step-Estimation-of-Impulsive-Time-Series} (clinical data experiments cannot be shared as the authors do not own the data).

\subsection{Synthetic data experiments}
Synthetic data experiments with first-order linear dynamics, defined by \eqref{eq:firstorder}, and second-order dynamics, defined by \eqref{eq:sys2}, are performed. The data generation is similar for both cases and is described in Appendix~\ref{app:synthdata}. The estimation is performed according to the descriptions above, with the parameter $\varepsilon$ set equal to the noise variance in the experiments with second-order dynamics, while a value of four times the noise variance is used in the first-order case.

\subsubsection{First-order dynamics} \label{sec:firstorderexp}
\begin{figure}
    \centering
    \expandafter\includegraphics\expandafter[\figsizepdf]{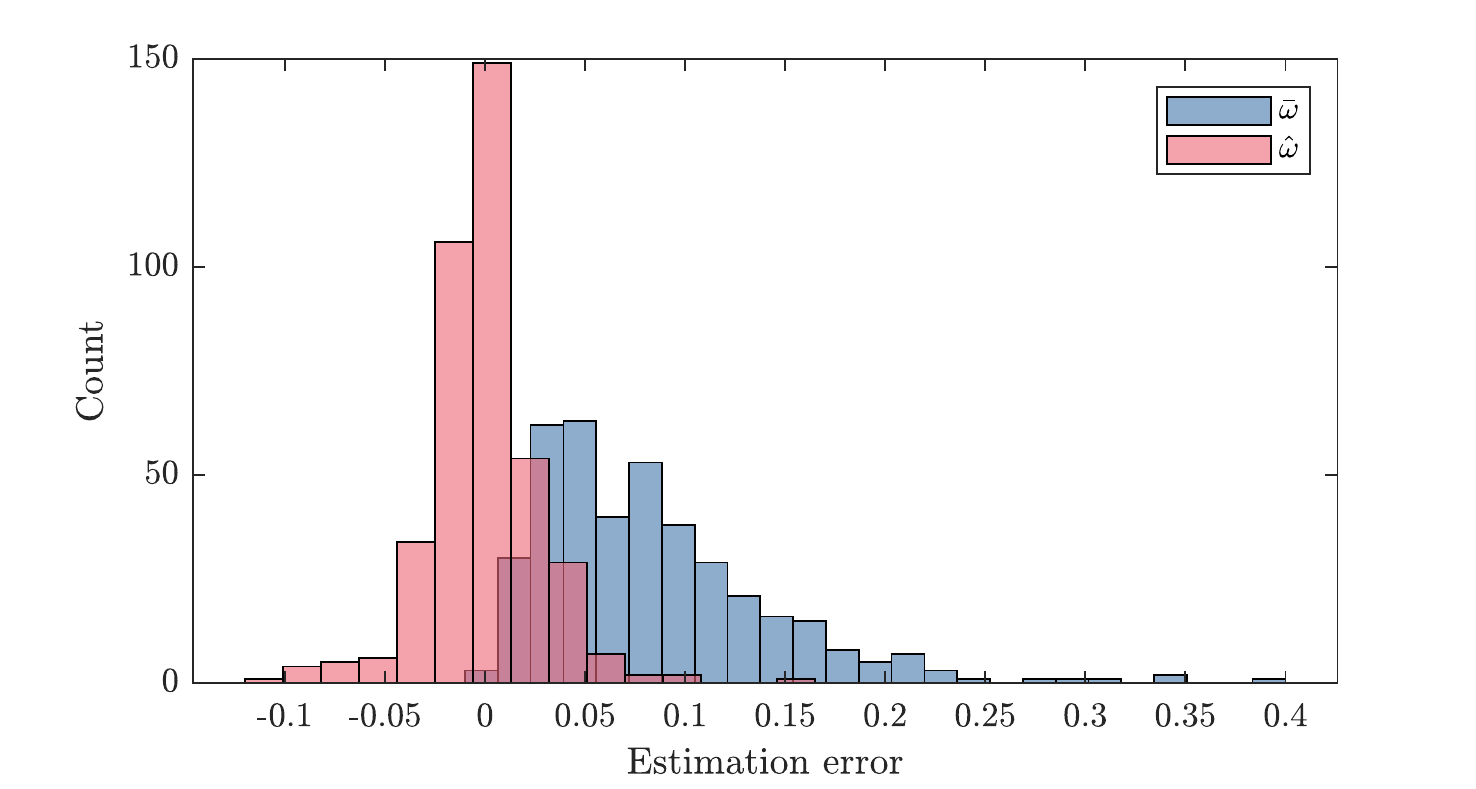}
    \caption{Histograms over $b-\bar \omega$ (initial estimation errors) and $b - \hat \omega$ (one-step estimation error) from $400$ Monte Carlo runs with synthetic data with first-order dynamics.}
    \label{fig:firstorderhist}
\end{figure}

\begin{figure}
    \centering
    \expandafter\includegraphics\expandafter[\figsizeeps]{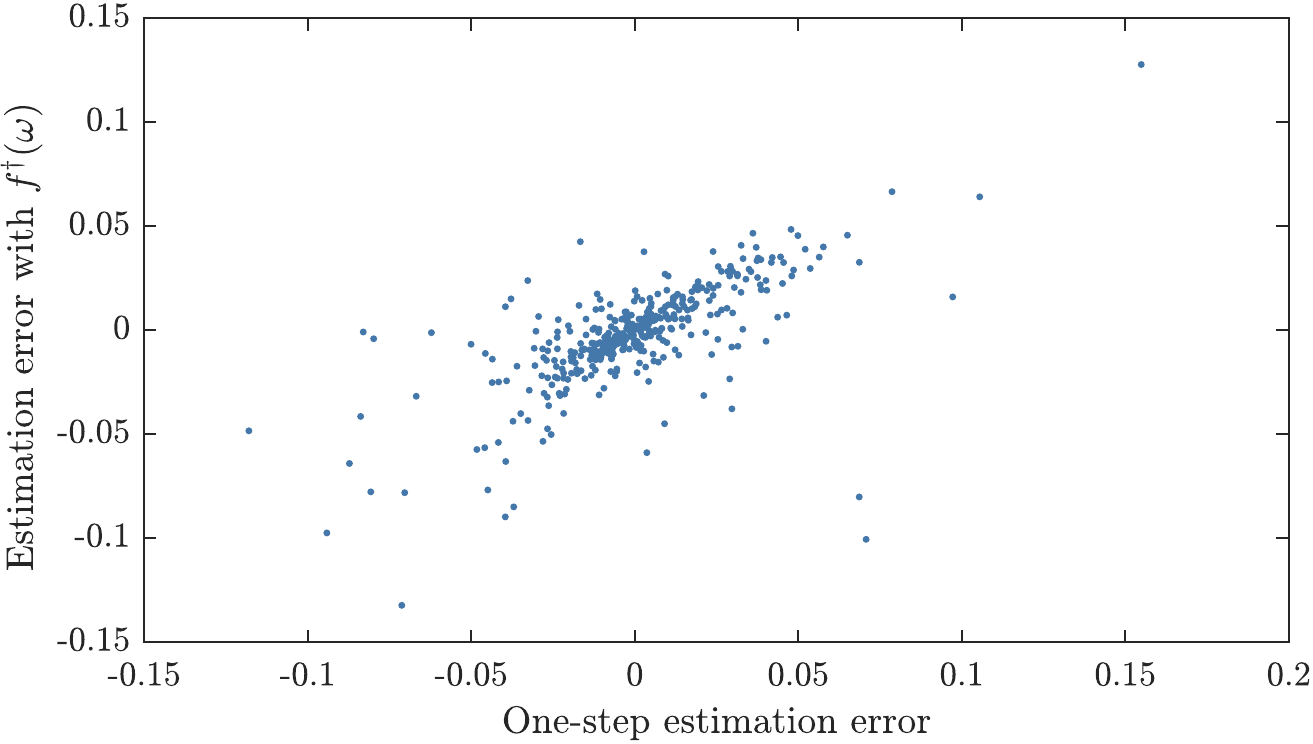}
    \caption{Time constant estimation errors for synthetic data with first-order dynamics. Errors when estimating with known impulse times are plotted against errors when time constant and elimination rate are estimated simultaneously with the one-step method.}
    \label{fig:firstorderfdaggercomp}
\end{figure}

Estimation according to \eqref{eq:omegahat} and \eqref{eq:omegabar2} is performed in $400$ Monte-Carlo runs. The resulting distributions of $\bar \omega$ and $\hat \omega$ are displayed in Fig.~\ref{fig:firstorderhist}; the improvement of the Newton step upon the preliminary estimate is clearly visible. In Fig.~\ref{fig:firstorderfdaggercomp}, the estimation errors of the one-step estimation method are plotted against the estimation error obtained when minimizing $f^\dagger(\omega)$ directly, i.e., when the impulse times are assumed to be known. The correlation coefficient is $0.68$, which indicates a strong correlation between the errors, and the increase in root-mean-square error when the impulse times are unknown is relatively small: $0.0278$ versus $0.0247$. This shows that the variance of the one-step estimate mainly comes from the the uncertainty of estimating the elimination rate with a known impulse time, and that \eqref{eq:omegabar2} is useful in determining the initial estimate. 

\subsubsection{Second-order dynamics: point estimates}\label{sec:secondorderexp}
Estimation of a second-order system with a fixed basal level is considered in this section. The estimates $(\hat b_1^{(n)},\hat b_2^{(n)})$ and $(\hat b_1^\mathrm{BIC},\hat b_2^\mathrm{BIC})$, calculated according to Section~\ref{sec:pointest} are compared against estimates $(\hat b_1^*,\hat b_2^*)$ obtained by minimizing $N_f$ over both $b_1$ and $b_2$, as suggested in \cite{rm22}. The evaluation is performed through a sequence of Monte-Carlo runs where the synthetic data are subject to noise with increasing variance. A typical data set is illustrated by the data subject to Gaussian i.i.d. noise in Fig.~\ref{fig:secondorderdata}.

\begin{table}[ht]
    \centering
    \caption{Estimation errors expressed as the mean Euclidean distance between the estimates and the true parameter values, and mean number of estimated impulses, for synthetic data subject to noise with standard deviation $\sigma$. Columns $(b_1^{(n)},b_2^{(n)})$ and $\gamma$ correspond to the estimates closest to the true parameters in the sets
    $\{(b_1^{(n)},b_2^{(n)})\mid n\in N\}$ and $n_0$ is the average number of estimated impulses.} 
    \label{tab:pointests}
    \begin{tabular}{c|ccccc}
        $\sigma$ & $(\hat b_1^{\mathrm{BIC}},\hat b_2^{\mathrm{BIC}})$ & $(\hat b_1^{(n)},\hat b_2^{(n)})$ & $(\hat b_1^*,\hat b_2^*)$ & $\gamma$ & $n_0$ \\
        \hline
        $0.002$ & $0.090$ & $0.030$ & $0.060$ & $0.0039$ & $4.01$ \\
        $0.004$ & $0.130$ & $0.062$ & $0.129$ & $0.0081$ & $3.84$ \\
        $0.006$ & $0.162$ & $0.083$ & $0.190$ & $0.0121$ & $3.76$\\
        $0.008$ & $0.185$ & $0.108$ & $0.269$ & $0.0159$ & $3.71$ \\
        $0.010$ & $0.218$ & $0.132$ & $0.349$ & $0.0202$ & $3.61$ \\
        $0.012$ & $0.245$ & $0.157$ & $0.423$ & $0.0250$ & $3.55$
    \end{tabular}
    
\end{table}

The estimation errors are summarized in Table~\ref{tab:pointests}. Regardless of the noise variance, the best of the Pareto-efficient estimates $(\hat b_1^{(n)},\hat b_2^{(n)})$ tends to be closer to the true parameter values than $(\hat b_1^*,\hat b_2^*)$. As the noise variance is increased, $(\hat b_1^{\mathrm{BIC}},\hat b_2^{\mathrm{BIC}})$ also starts to outperform $(\hat b_1^*,\hat b_2^*)$. Note also that $\gamma$ is much closer to the true parameters than any point estimate, i.e. the estimation errors are mostly caused by choosing a wrong point along $\gamma$.

Generally, the number of estimated impulses exceeds the true number of impulses, but curiously, the estimate improves when the noise level increases. The use of the BIC with many parameters compared to the amount of data is a potential cause. However, it should be noted that our method makes no assumptions about the weight of the impulses as the regularization is done implicitly. For synthetic data sets, incorporating information regarding the impulse weights in the estimation could potentially improve the performance, but for clinical data, making similar assumptions about the magnitude of secretion events may be unwarranted.

\subsubsection{Second-order dynamics: basal level}\label{sec:basalexp}
\begin{figure}
    \centering
    \expandafter\includegraphics\expandafter[\figsizepdf]{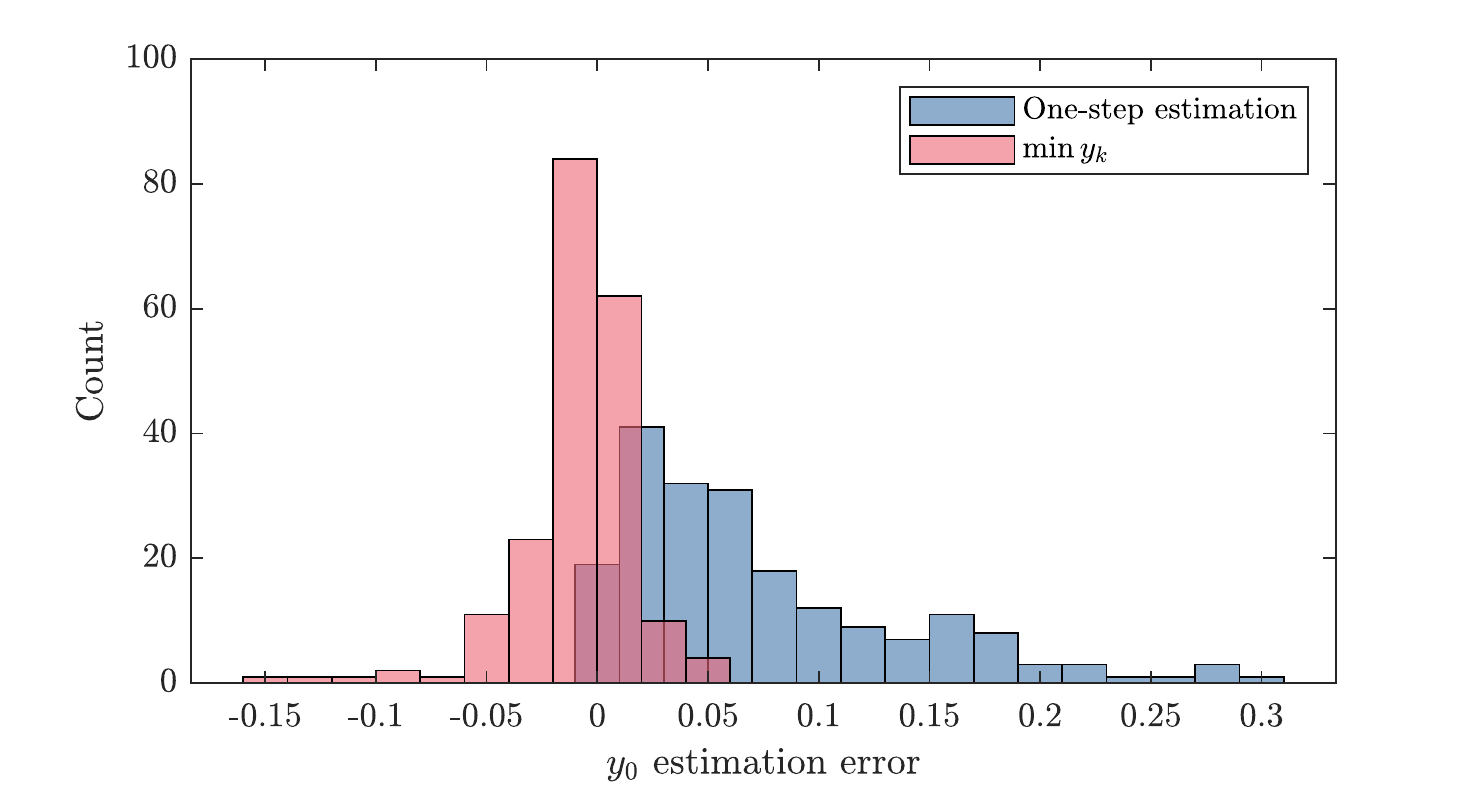}
    \caption{Histograms over one-step estimation error of basal level, and minimum value of output from $200$ Markov runs with synthetic data.}
    \label{fig:basalhist}
\end{figure}

Synthetic data where one elimination rate ($b_2$) is significantly slower than the other is used to evaluate the estimation of the basal level $y_0$, and the parameters $b_1,b_2$, according to the description in Section~\ref{sec:basal}. Histograms of the basal level estimation error and the minimal value (compared to the basal level) of the sampled data from $200$ Monte-Carlo runs are displayed in Fig.~\ref{fig:basalhist}. It can be seen that estimating the basal level outperforms the simplistic approach of choosing the minimal value as the basal level, and they display biases in opposite directions. The bias of the latter method is expected since the impulse response of the system only approaches the basal level as time tends to infinity, while the bias of former is discussed further below. In Table~\ref{tab:basalest}, the performance of the estimation of basal level and 
elimination rates are given. As expected, the estimation performance for the slower elimination rate is significantly better than for the faster.

\begin{table}[ht]
    \centering
    \caption{Elimination rate and basal level estimation performance from synthetic data experiment.} 
    \label{tab:basalest}
    \begin{tabular}{c|ccc}
        Estimate & Bias & Variance \\
        \hline
        $b_1$ & $0.327$ & $1.62$ \\
        $b_2$ & $0.00462$ & $0.00456$ \\
        $y_0$, one-step est. & $-0.00949$ & \num{6.73e-4} \\
        $y_0$, $\min y_k$ & $0.0739$ & $0.00425$ \\
    \end{tabular}
\end{table}

\begin{figure}
    \centering
    \expandafter\includegraphics\expandafter[\figsizeeps]{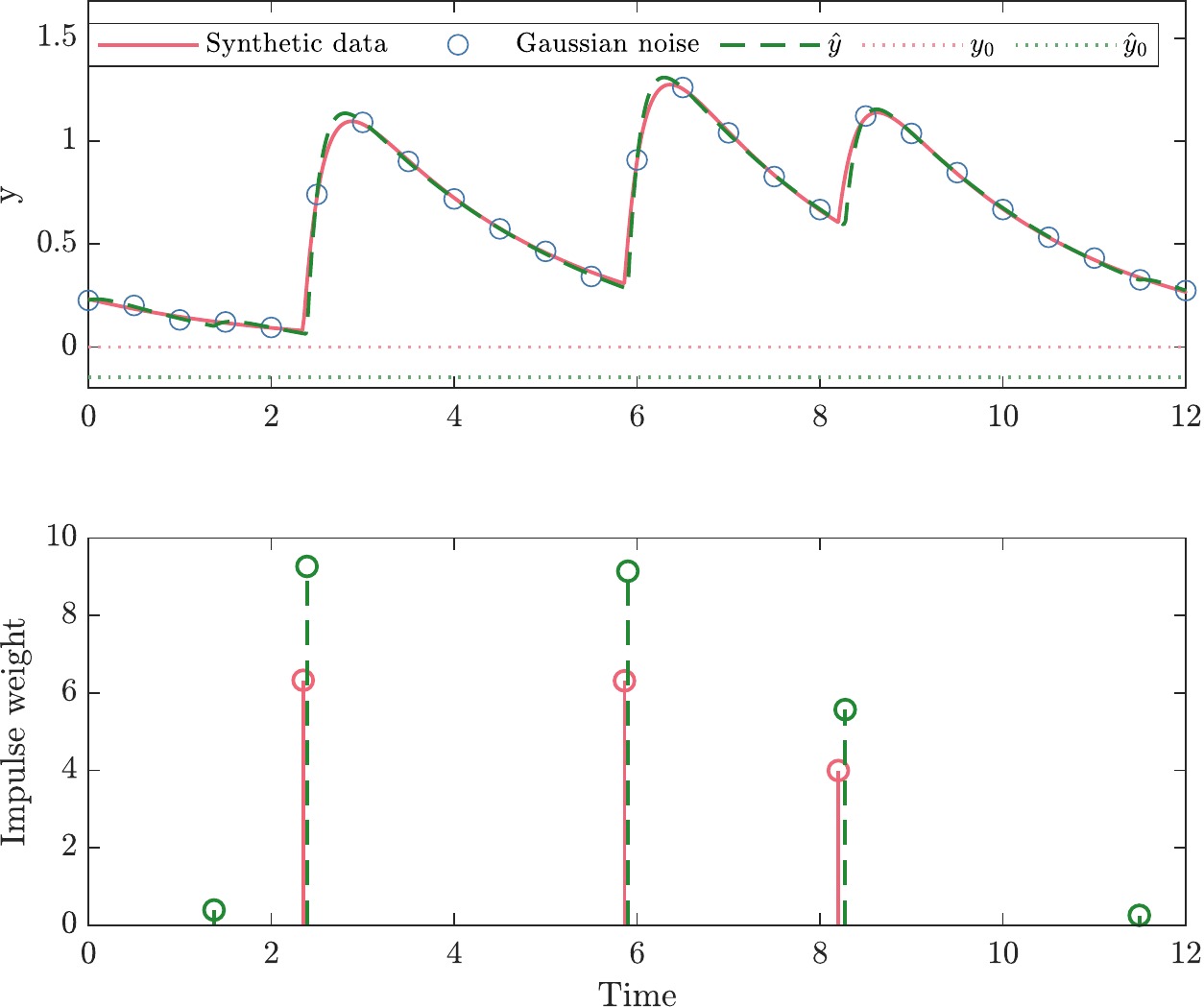}
    \caption{Top: Synthetic data and simulated output from estimated model with large basal level estimation error. Bottom: Corresponding synthetic and estimated impulses.}
    \label{fig:basalex}
\end{figure}

Fig.~\ref{fig:basalhist} indicates that a few data sets give rise to a significant negative basal level estimation error. The synthetic and estimated parameters of the most extreme case are summarized in Table~\ref{tab:basalex} and the corresponding output is illustrated in Figure~\ref{fig:basalex}. There are clear discrepancies for all parameters, however the residual sum is lower with the estimated parameters, so the estimator apparently finds alternative solutions where additional impulses yield a better fit than for the noisy original data. Similar results can be seen for other data sets where the estimation errors are large. Imposing stricter restrictions on the number of impulse estimates would prevent these problems, but we chose to retain them in order to keep the estimation assumptions at a minimum, and to illustrate the challenging nature of this estimation problem.
\begin{table}[ht]
    \centering
    \caption{True and estimated parameter values from synthetic data with large basal level estimation error.} 
    \label{tab:basalex}
    \begin{tabular}{c|ccccc}
         & $b_1$ & $b_2$ & $y_0$ & $n_0$ & Residual error \\
        \hline
        Data & $4.76$ & $0.460$& $0$ & $3$ & \num{1.88e-3} \\        
        Estimate & $7.14$ & $0.371$ & $-0.148$ & $5$ & \num{6.43e-4}
    \end{tabular}
\end{table}

\subsubsection{Second-order dynamics: robustness against outliers}\label{sec:robustexp}
\begin{figure}
    \centering
    \expandafter\includegraphics\expandafter[\figsizeeps]{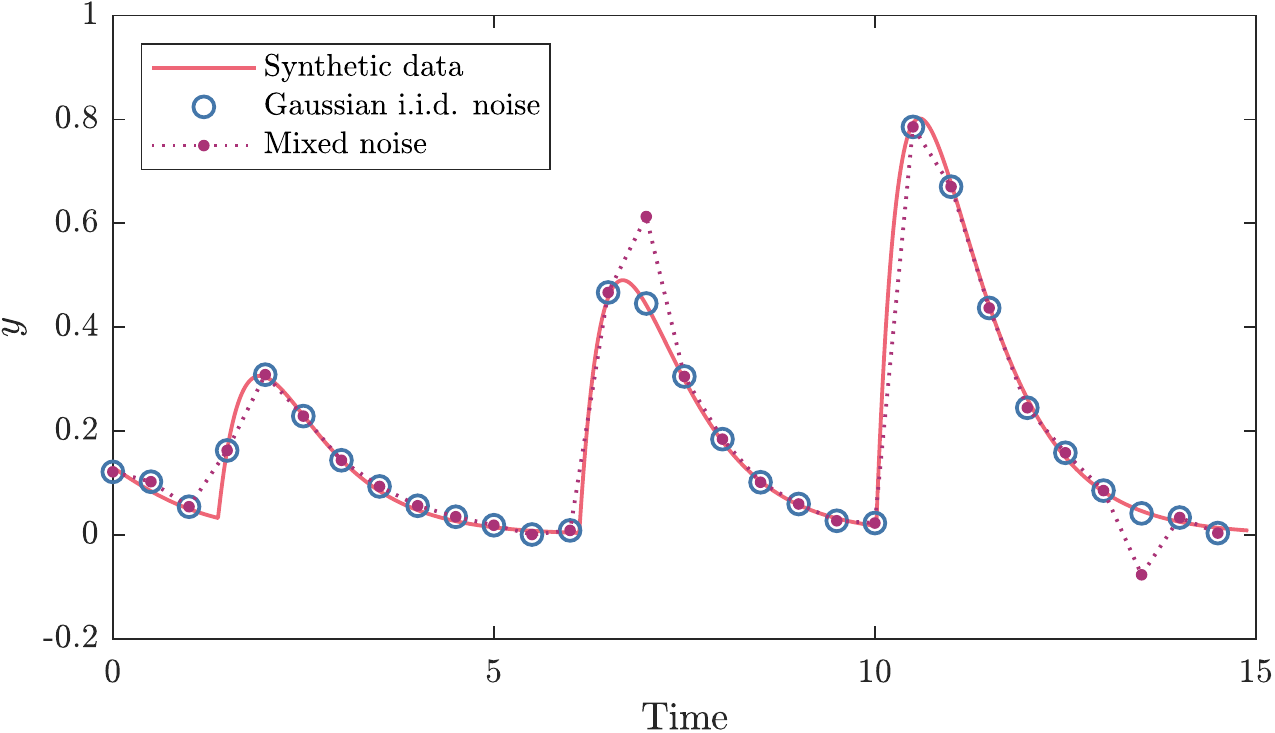}
    \caption{Example of a synthetic data set, subject to Gaussian i.i.d noise, and a mixture of Gaussian noise and uniformly distributed noise with higher variance.}
    \label{fig:secondorderdata}
\end{figure}

To evaluate the impact of outlying measurements on the estimation, synthetic data subject to two different types of noise are used. In the base case, Gaussian i.i.d. noise is applied to all data points. In an alternative setup, the aforementioned noise is applied to all but two data points, which instead are subject to uniformly distributed noise with significantly higher variance. In Fig.~\ref{fig:secondorderdata} one data set subject to the two noise types is illustrated.

\begin{figure}
    \centering
    \expandafter\includegraphics\expandafter[\figsizeeps]{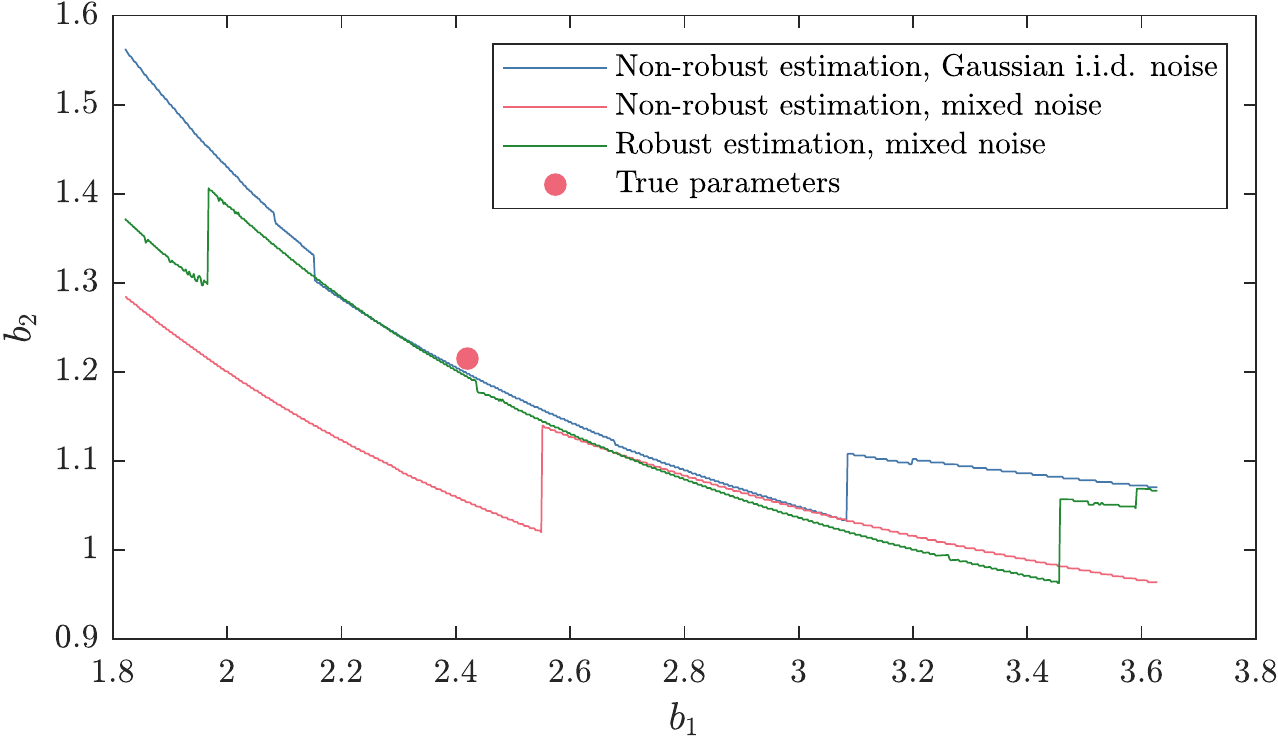}
    \caption{Non-robustly and robustly estimated $\gamma$-curve from the synthetic data sets in Fig.~\ref{fig:secondorderdata}.}
    \label{fig:secondorderrobust}
\end{figure}

The estimation method is evaluated in $50$ Monte-Carlo runs, where the standard one-step estimation is performed for both the i.i.d.~and mixed noise cases, while robust estimation is only performed for the mixed noise case. The threshold of $\epsilon=0.1$ is used, which corresponds to reducing the effective sample size by approximately two (depending on the size of the individual data sets). Estimated $\gamma$-curves for these cases, based on the data in Fig.~\ref{fig:secondorderdata}, are displayed in Fig.~\ref{fig:secondorderrobust}. The distance between the curves and the true parameter values indicate that the outliers clearly deteriorate the performance of the non-robust estimation, however, when the robust algorithm is used, the performance is recovered almost fully. The discontinuities that can be observed for all three curves are caused by different local minima corresponding to to the global minimum. Local minima are typically associated with different sets of nonzero impulse estimates, which is briefly discussed in Appendix~\ref{app:firstordersens}.

In Table~\ref{tab:robustests}, an evaluation of both the estimated curves and point estimates are provided for the Monte-Carlo runs. The small difference between the results for the base case and the robust estimation, and the large deviation when the standard method is used on mixed noise, show the usefulness of the robust method. Also note that, in the latter case, the minimizer of $N_f$ at times coincides with the minimal value of $I_{b_1}$, so if a larger parameter range was used in the estimation, the performance could deteriorate even further.


\begin{table}[ht]
    \centering
    \caption{Mean Euclidean distance between estimated $\gamma$-curve and point estimates $(\hat b_1^{\mathrm{BIC}},\hat b_2^{\mathrm{BIC}})$ from synthetic data experiments. Point estimates for the mixed noise case with standard estimator are not included as the large error in $\gamma$ renders the estimates useless.} 
    \label{tab:robustests}
    \begin{tabular}{c|cc}
        Setup & $(\hat b_1^{\mathrm{BIC}},\hat b_2^{\mathrm{BIC}})$ & $\gamma$ \\
        \hline
        Gaussian i.i.d. noise, standard estimation & $0.173$ & $0.0134$  \\
        Mixed noise, standard estimation &-& $0.0739$ \\
        Mixed noise, robust estimation, mixed noise & $0.191$ & $0.0160$  \\       
    \end{tabular}
\end{table}

\subsection{LH data experiments}
The one-step estimation method is now used on clinical data. We use a data set with LH blood concentrations collected from healthy males, which was collected in experiments described in \cite{ltr05}. A more rigorous analysis of this data, including the effect of a selective gonadotropin releasing hormone receptor antagonist, is included in \cite{rm22b}. The focus here is instead on the features of the algorithm introduced in this work, which are exemplified on individual data sets. For these experiments, we note that the elimination rates for LH and GnRH are expected to satisfy
\begin{equation} \label{eq:bbounds}
    0.23~\si{min^{-1}}\le b_1<0.69~\si{min^{-1}},\quad 0.0087~\si{min^{-1}} <b_2\le 0.014~\si{min^{-1}},
\end{equation}
according to \cite{kv98}.

\subsubsection{Robust basal level estimation}
\begin{figure}
    \centering
    \expandafter\includegraphics\expandafter[\figsizeeps]{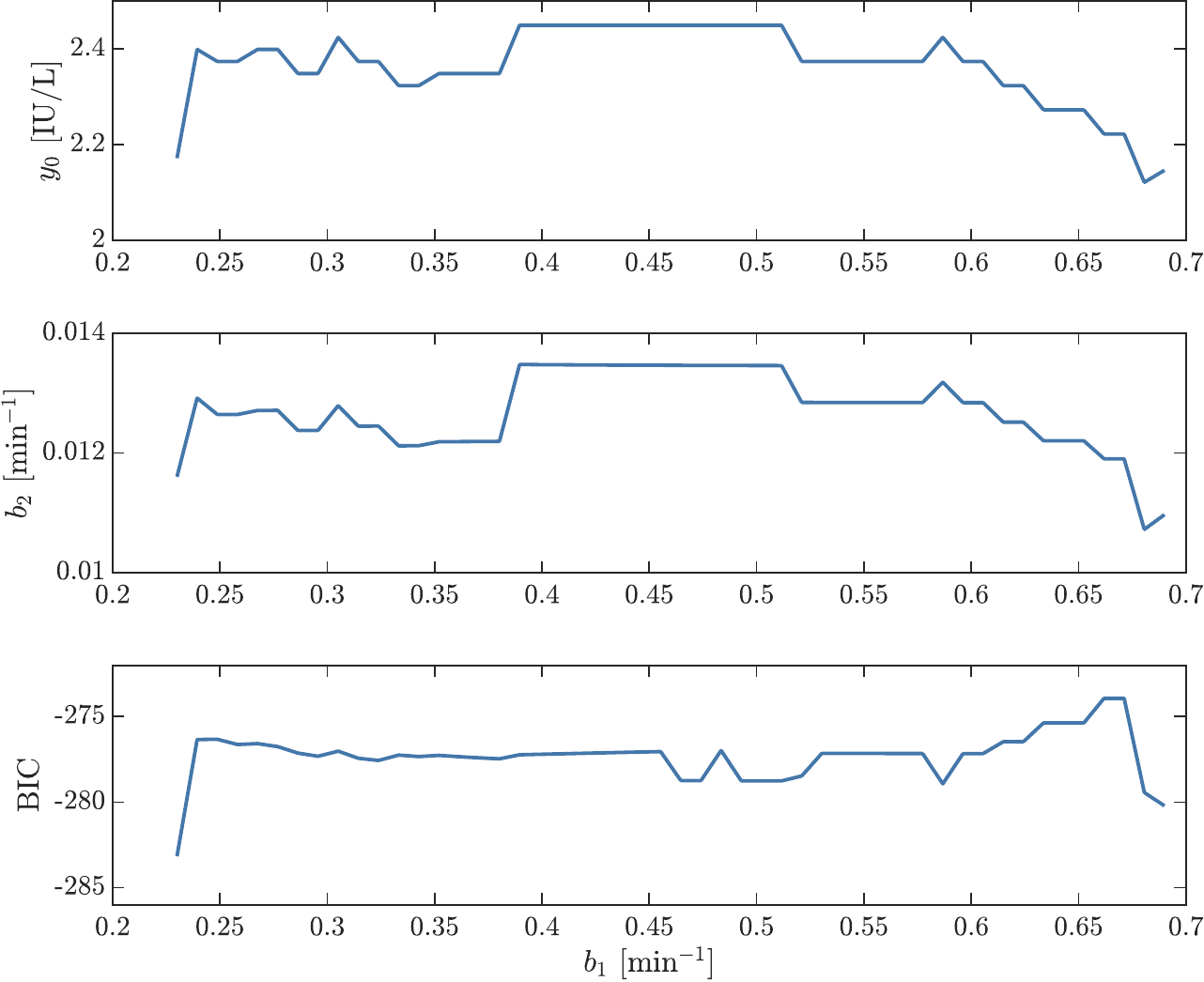}
    \caption{Estimated basal level (top), LH elimination rate (middle) and BIC (bottom) depending on GnRH elimination rate for LH data shown in Fig.~\ref{fig:LHsim}.}
    \label{fig:LHest}
\end{figure}

\begin{figure}
    \centering
    \expandafter\includegraphics\expandafter[\figsizeeps]{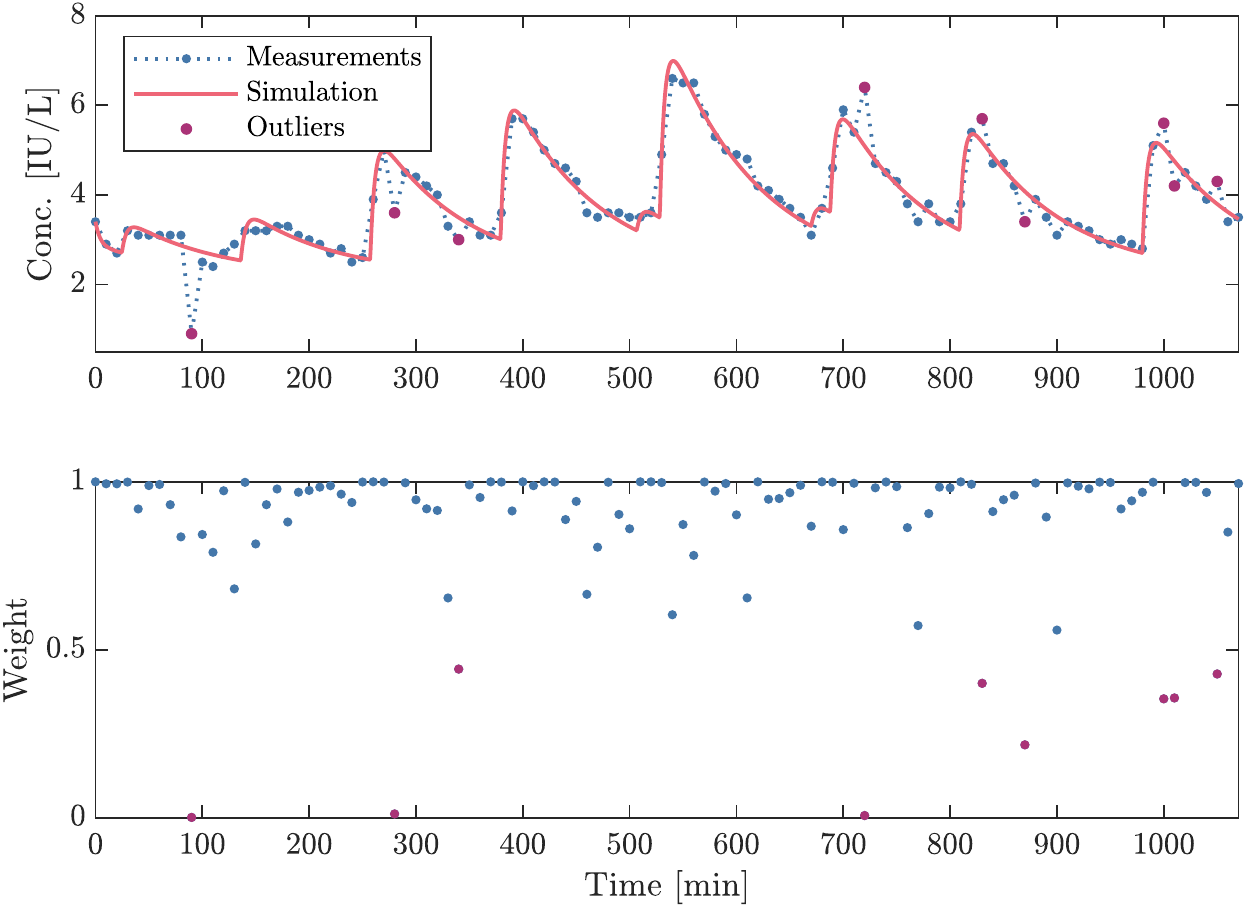}
    \caption{Top: LH measurements from a $32$-year old male, and simulated output from estimated system. Bottom: weighting of data points from robust least squares solver. Points weighted below $0.5$ are highlighted in red in both plots.}
    \label{fig:LHsim}
\end{figure}

Estimation of the basal level, the elimination rates, and the secretion events is performed on hormone data of a $32$-year old healthy male. The data set consists of $108$ measurements of LH sampled every ten minutes. Since large measurement errors are suspected for several data points, robust estimation is needed. The methods described in Section~\ref{sec:basal} and Section~\ref{sec:robust} are therefore combined, with the range of $b_1$ values given by \eqref{eq:bbounds}, and the parameter $\epsilon$ set to $5/108$, i.e. an effective sample size of $103$ is assumed. The estimated values of $b_2$ and $y_0$, and the BIC, are plotted against $b_1$ in Fig.~\ref{fig:LHest}. 
As expected, the sensitivity of the estimates to $b_1$  is relatively low, but curiously the lowest BIC score appear at both edges of the parameter range. Also note that the estimated values of $b_2$ are biologically viable and satisfy \eqref{eq:bbounds}. The simulated response of the estimated model corresponding to the lowest BIC score is illustrated in Fig.~\ref{fig:LHsim}. There, the weights of the data points obtained from the robust estimator are also displayed and the most down-weighted points are highlighted.

\subsubsection{Outlying hormone profile}
\begin{figure}
    \centering
    \expandafter\includegraphics\expandafter[\figsizeeps]{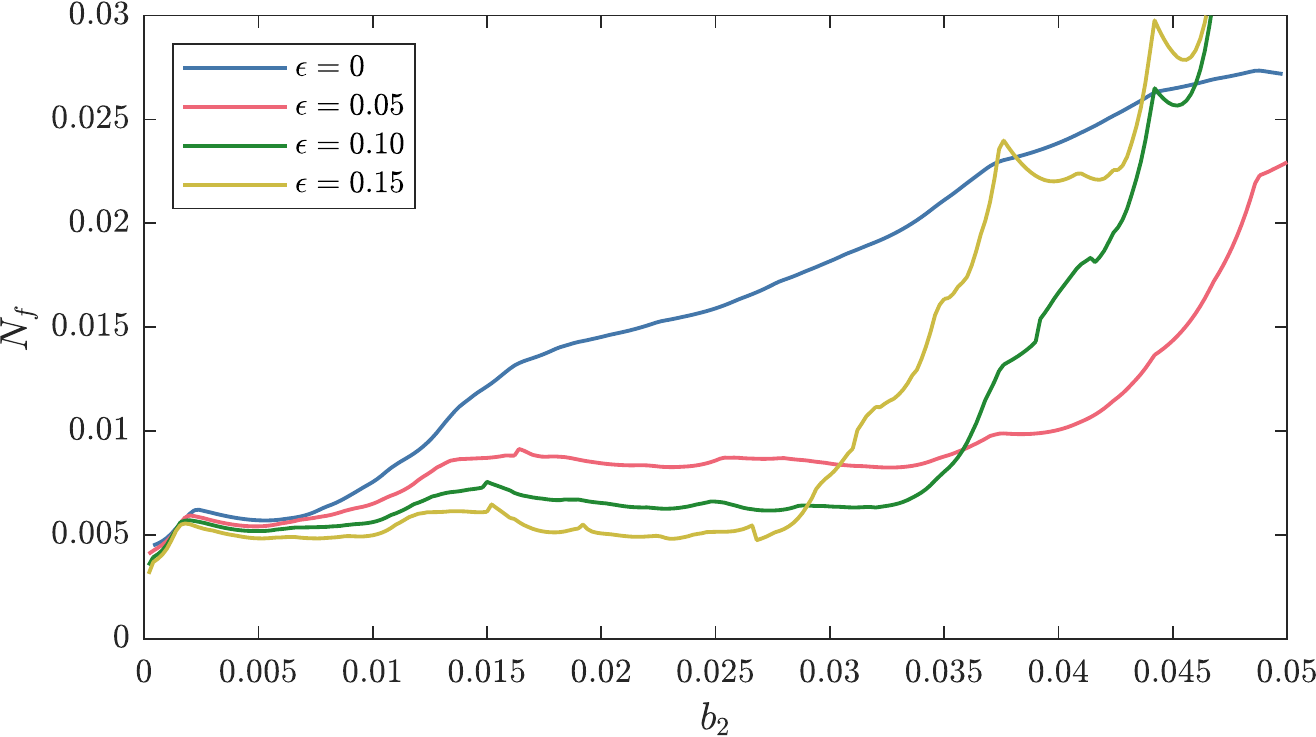}
    \caption{$N_f$-curves for standard ($\epsilon=0$) and robust estimation ($\epsilon>0$) for LH data shown in Fig.~\ref{fig:LHoutliersim}. }
    \label{fig:LHNf}
\end{figure}

\begin{figure}
    \centering
    \expandafter\includegraphics\expandafter[\figsizeeps]{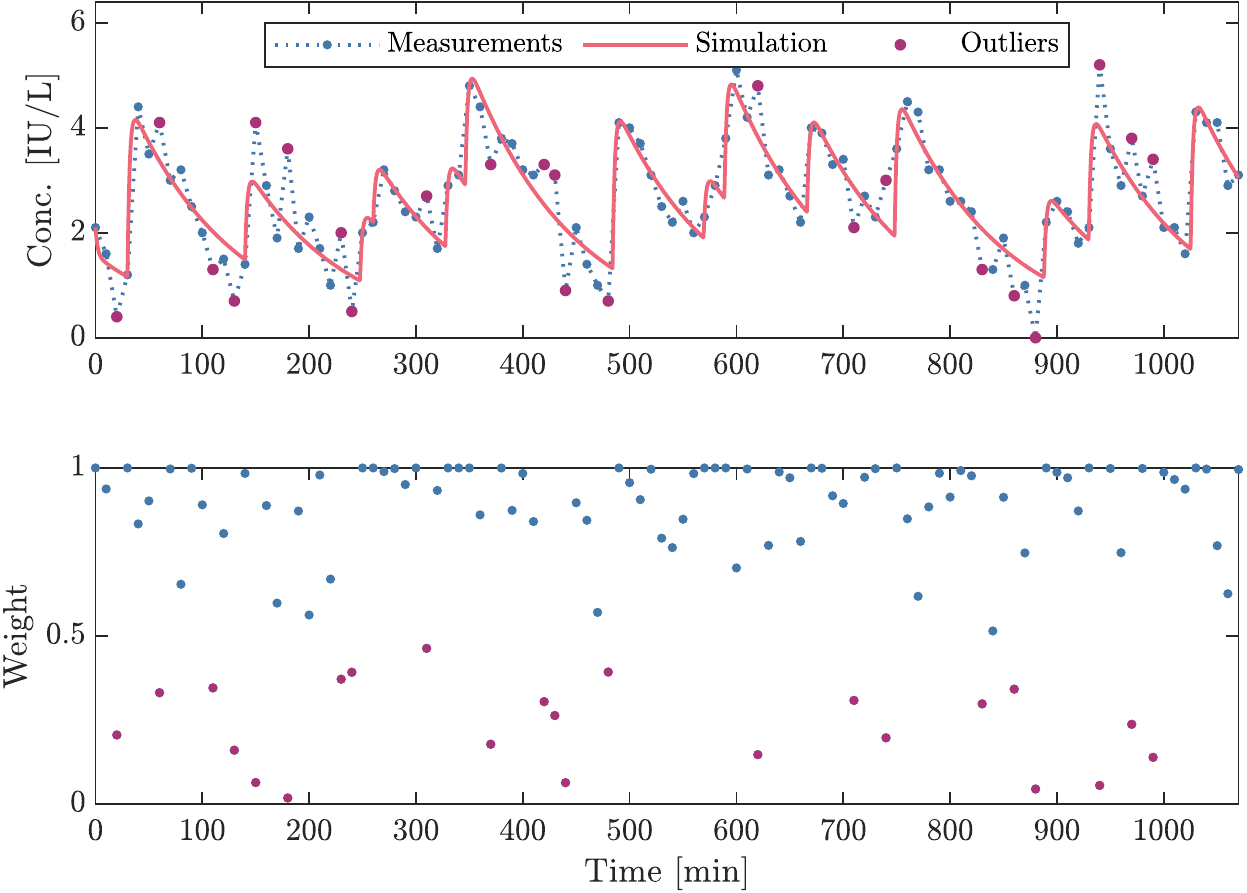}
    \caption{Top: LH measurements from a $68$-year old healthy male, and simulated output from estimated model. Bottom: weighting of data points from robust least squares solver. Points weighted below $0.5$ are highlighted in red in both plots.}
    \label{fig:LHoutliersim}
\end{figure}

The hormone profile of a $68$-year old healthy male that appears inconsistent with the assumed model is now analyzed. For simplicity, the basal level is assumed to coincide with the lowest measured LH concentration and $b_1=0.5$ is fixed; very similar results are obtained with other parameter values. Standard and robust estimations of $b_2$, with different values of $\epsilon$, are performed. The corresponding functions $N_f$ are displayed in Fig.~\ref{fig:LHNf}. Three observations can be made from this plot. First, the effect of the robust estimation is to decrease $N_f$ for larger values of $b_2$. This behavior is generally seen for data sets with outliers, and is the same mechanism by which robust estimation moves the $\gamma$-curve closer to the true parameters in the presence of outliers in Fig.~\ref{fig:secondorderrobust}. Second, depending on which interval $I_{b_2}$  and value of $\epsilon$ that are chosen, local minima of $N_f$ could be identified as solutions to \eqref{eq:b2bar}. The simulated output for such an estimate of the system with $I_{b_2}= [0.003, 0.01]$ and $\epsilon=0.01$ is shown together with the weights of the robust estimator in Fig.~\ref{fig:LHoutliersim}. Third, the global minimizer of $N_f$ is in fact $b_2=0$ for all $\epsilon$, which indicates that the data are inconsistent with the assumed model according to the discussion in Section~\ref{sec:outlyingind}. The reliability of estimates such as the one illustrated in Fig.~\ref{fig:LHoutliersim} is therefore questionable, something that is also indicated by the large number of outliers. It is known that GnRH pulses appear at higher frequency in older males, so increasing the sampling rate of the measurements is probably a better strategy to recognize the impulsive events for this individual.


\section{Discussion and conclusions}\label{sec:concl}
There are many possible approaches to analyzing hormone concentration time-series data. The aim of the present work has been to develop a method involving minimal assumptions or manual tuning, which also is well-motivated mathematically. However, achieving these goals simultaneously is challenging, particularly when the algorithm is adapted for clinical data. We have here only attempted a theoretical analysis of the first-order case, but as the model and data depart from this situation, the algorithm becomes more involved and less tractable.

An advantage of the presented method is that estimates with different resolutions can be obtained. Keeping the uncertainties in model and measurements and the inherent ill-posedness of the estimation problem in mind, the reliability of any point estimate from clinical data is probably low. All estimates produced along a section of the curve $\gamma$ might therefore be a better representation of the range of possible parameters and secretion events of a given data set. Such lower resolution estimates have the additional advantage of a more transparent mathematical derivation.

Possible future research directions include the application of the one-step estimation method under other modeling assumption. Generalization of the method has already been presented in this work, in the form of applying it to estimate the basal level and incorporating a robust estimator. This indicates that other generalizations may also be possible.

\bibliography{refs}

\appendix
\section{Sensitivity of estimation of a first-order system}\label{app:firstordersens}
We wish to analyze the estimation performance from the response of system \eqref{eq:firstorder} to a series of impulses. But, as the response of each impulse can be viewed as a separate least squares problem, the analysis is restricted to consider a single impulse with amplitude $d$ at $t=0$. By summing over all impulses, corresponding formulas for the general case are obtained.

\subsection{Residual sum for impulse response}
Consider first the noise-free impulse response. If the impulse time is known, the residual sum of squares is given by
\begin{equation*}
    f^\dagger_0(\omega) = \sum_{k=1}^K (\alpha_0(\omega) \e^{-\omega t_k} - d \e^{-bt_k})^2,
\end{equation*}
where the impulse weight $\alpha_0(\omega)$ which minimizes the squared error is given by
\begin{equation*}
    \alpha_0(\omega)= d~\frac{ \sum\limits_{k=1}^K \e^{-(\omega+b)t_k}}{\sum\limits_{k=1}^K \e^{-2\omega t_k}}.
\end{equation*}
If the measurements are corrupted by additive zero-mean i.i.d. noise with (finite) variance $\sigma^2$, i.e. $y(t_k) = x(t_k) + \epsilon_k$, the corresponding residual sum becomes
\begin{multline*}
    f^\dagger(\omega) = f^\dagger_0(\omega) + \sum_{k=1}^K (2\eta \alpha_0(\omega) + \eta^2)\e^{-2bt_k}\\
    + \sum_{k=1}^K(2d \e^{bt_k}\epsilon_k + \epsilon_k^2 - 2(\alpha_0(\omega) \e^{-\omega t_k} \epsilon_k + \eta \e^{-\omega t_k} (d \e^{-\omega^* t_k}+\epsilon_k))),
\end{multline*}
where
\begin{equation*}
    \alpha(\omega) = \alpha_0(\omega)+ \eta(\omega), \quad \eta(\omega) = \frac{\sum\limits_{k=1}^K \e^{-bt_k}\epsilon_k}{\sum\limits_{k=1}^K \e^{-2bt_k}}.
\end{equation*}
The expected value of $f^\dagger(\omega)$ is given by
\begin{equation*}
    \mathbb{E} f^\dagger(\omega) = f^\dagger_0(\omega)+c_0,
\end{equation*}
where $c_0=(K-1)\sigma^2$. In the following, we suppress the argument of $\alpha_0$, $\alpha$, $\eta$ and their derivatives.

The estimation is based on approximating $f^\dagger(\omega)$ as a quadratic function. In the noise-free case, the derivatives are
\begin{equation*}
    \frac{\partial f_0^\dagger}{\partial \omega} = 2 \sum_{k=1}^K (\alpha_0 \e^{-\omega t_k} - d \e^{-bt_k}) \bigg(\frac{\partial \alpha_0}{\partial \omega}-t_k \alpha_0\bigg)\e^{-bt_k},
\end{equation*}
\begin{multline*}
    \frac{\partial^2 f^\dagger_0}{\partial \omega^2}\\
    = 2 \sum_{k=1}^K \bigg(\bigg(\frac{\partial \alpha_0}{\partial \omega}-t_k \alpha_0\bigg)^2\e^{-2bt_k} + (\alpha_0 \e^{-\omega t_k} - d \e^{-bt_k})\bigg(\frac{\partial^2 \alpha_0}{\partial \omega^2} - 2 t_k \frac{\partial \alpha_0}{\partial \omega} + t_k^2 \alpha_0 \bigg) \bigg)\e^{-bt_k}.
\end{multline*}
When evaluated at the minimum $\omega=b$, $\alpha_0=d$, which leads to
\begin{equation*}
    \frac{\partial f^\dagger_0}{\partial \omega}\bigg|_{\omega=b} = 0,\quad \frac{\partial^2 f^\dagger_0}{\partial \omega^2}\bigg|_{\omega=b} = 2 c_2,
\end{equation*}
where 
\begin{equation*}
    c_2 = \sum_{k=1}^K \bigg(\frac{\partial \alpha_0}{\partial \omega}\Big|_{\omega=b}-t_k d\bigg)^2\e^{-2bt_k},
\end{equation*}
so
\begin{equation*}
    f^\dagger_0(\omega) = c_2 (\omega - b)^2  + O((\omega-b)^3).
\end{equation*}
By dominated convergence and boundedness of moments of the noise, expectations and derivatives can be interchanged, so
\begin{equation*}
    \mathbb E f^\dagger(\omega) = c_0 + c_2(\omega - b)^2 + O((\omega-b)^3).
\end{equation*}
In this construction, $c_2 \sigma^{-2}$ can be identified as the Fisher information for the parameter $b$.

\subsection{Multiple impulse estimates}
The formula in \eqref{eq:omegatilde}, which approximates the transition between one and multiple impulses being estimated from a noisy impulse response, is derived here. Due to the challenging combinatorial nature of the problem, a number of approximations are made.

As an auxiliary step, note that the sum that defines $c_2$ also can be interpreted using a discrete random variable $X$, given by
\begin{equation*}
    P(X=t_k) = \frac{\e^{-2bt_k}}{\sum_{k=1}^K \e^{-2bt_k}}, \quad k=1,\dots K.
\end{equation*}
The mean and variance of $X$ then satisfy
\begin{align*}
    d~ \mathbb E X = \frac{\partial \alpha_0}{\partial \omega}\Big|_{\omega=b},\\
    d^2~\mathrm{Var} X = \sum_{k=1}^K \bigg(\frac{\partial \alpha_0}{\partial \omega}\Big|_{\omega=b}-t_k d\bigg)^2\e^{-2bt_k}\bigg(\sum\limits_{k=1}^K \e^{-2bt_k}\bigg)^{-1} = c_2 \bigg(\sum\limits_{k=1}^K \e^{-2bt_k}\bigg)^{-1}.
\end{align*}

Now consider the transition between one (located at time $t_1$) and two (at $t_1$ and $t_{m+1}$) nonzero impulse estimates. The setup with two nonzero impulses can be viewed as two separate least squares estimation problems so the total residual sum of squares is the sum of the residuals from the two. That implies that $c_0$ decreases from $(K-1)\sigma^2$ to $(K-2)\sigma^2$ when transitioning from one to two impulses. Now consider the interpretation of $c_2$ as 
\begin{equation*}
    c_2 = d^2\sum\limits_{k=1}^K \e^{-2bt_k}~\mathrm{Var} X,
\end{equation*}
in the single impulse case. With two impulses, the corresponding parameter $c_{2,2}$ is given by
\begin{equation*}
    c_{2,2} = d^2\sum\limits_{k=1}^m \e^{-2bt_k}~\mathrm{Var} X_1 + d^2\sum\limits_{k=m+1}^K \e^{-2bt_k}~\mathrm{Var} X_2,
\end{equation*}
where $X_1$ and $X_2$ are defined analogously to $X$ above. It is not hard to see that $c_{2,2}\le c_2$, however the degree reduction depends on $m$. The extreme case $m=2$ leads to $\mathrm{Var} X_1=0$ and $\mathrm{Var} X_2\approx \mathrm{Var} X$, under the assumption that $X$ is well-approximated by a geometric distribution. That leads to
\begin{equation*}
    \frac{c_{2,2}}{c_2} \approx \frac{\sum\limits_{k=2}^K \e^{-2bt_k}}{\sum\limits_{k=1}^K \e^{-2bt_k}} \le \frac{K-1}{K}.
\end{equation*}
On the other hand, if $m=K$, the effect on either the sums or the variances is negligible, so in particular we have
\begin{equation*}
    \frac{c_{2,2}}{c_2} \ge \frac{K-2}{K-1}.
\end{equation*}
We can now conclude that $\min_\omega N_f(\omega)$, which is approximated by $\sqrt{c_0/c_2}$, may either increase or decrease when two impulses are estimated rather than one and that the effect will depend on the location of the second impulse. More general transitions of this kind are changes between different sets of nonzero impulse estimates, which can result in different local minima for $N_f$.

To estimate $\tilde \omega$, we consider only the case $m=1$ and estimate the probability of $d_2$ being nonzero. Assume $t_1=0$, $\omega\le b$ and that the estimated state of the system satisfies
\begin{equation}\label{eq:stateEst}
    \hat x(t_k) = \alpha_0 \e^{-bt_k}
\end{equation}
for $k\ge 2$, i.e.~$\alpha(b)$ is approximated by its expected value $\alpha_0(b)$, when only the indices $k=2,3,\dots K$ are considered. This results in two cases depending on the first noise term $\epsilon_1$:
\begin{itemize}
    \item If $\epsilon_1\ge\alpha_0(\omega)-d$, $d_2$ is estimated to be zero to minimize the residual error at $t_1$;
    \item If $\epsilon_1<\alpha_0(\omega)-d$, $d_1$ can be chosen to give zero error at $t_1$, while a positive $d_2$ can be chosen to keep \eqref{eq:stateEst} satisfied. 
\end{itemize}
We are therefore interested in the probability
\begin{equation*}
    P(\epsilon_k)\ge \alpha_0(\omega)-d,
\end{equation*}
i.e.~the probability of one impulse being used rather than two, and for which $b$ this probability is significantly larger than zero. Assuming Gaussian noise, the probability distribution is linearized around $\epsilon_k=0$, i.e. $\omega=b$, and the value $\bar \omega$ where zero is crossed is derived. That gives
\begin{equation*}
    b - \bar \omega = \frac{\sigma}{\partial \alpha_0/\partial \omega} \sqrt{\pi/2}
\end{equation*}
Now assume equidistant sampling, so $t_{k+1}-t_k = T$, and approximate $X$ as a geometric distribution, scaled to take values $kT, k=0,1,\dots$. The expected value and variance of $X$ then satisfy the relation
\begin{equation*}
    \mathrm{Var} X \approx (\mathbb E X)^2 \e^{2b T},
\end{equation*}
which implies
\begin{equation*}
    \frac{\partial \alpha_0}{\partial \omega} \approx \e^{-bT} \sqrt{c_2(1-\e^{-2bT})},
\end{equation*}
which in turn leads to \eqref{eq:omegatilde}.

\subsection{Deviations from quadratic approximation}
We consider the residual sum $f_{c_3}$ on the form \eqref{eq:cubicf} and investigate the sensitivity with respect to $c_3$ of the estimation method. The analysis is done through linearization and includes the terms $\bar \omega$ and $N_{f_{c_3}}(\bar \omega)$. The first is characterized by the derivative being zero, which implies 
\begin{equation*}
    \bigg(\frac{\partial f}{\partial \omega}\Big|_{\omega=\bar \omega}\bigg)^2 = \Big(f(\omega) \frac{\partial^2 f}{\partial \omega^2}\Big)\Big|_{\omega=\bar \omega},
\end{equation*}
which yields
\begin{equation*}
    2c_0c_2+6c_0c_3z-2c_2^2z^2-4c_2c_3z^3-3c_3^2z^4=0,
\end{equation*}
where $z=\bar \omega-b$. Differentiating with respect to $c_3$ and solving for the derivative results in the sensitivity
\begin{equation*}
    \frac{\partial z}{\partial c_3}\Big|_{c_3=0} = \frac{c_0}{2c_2^2}.
\end{equation*}
The sensitivity of the second term is calculated similarly and yields
\begin{equation*}
    \frac{d N_{f_{c_3}}(\bar \omega)}{d c_3}\Big|_{c_3=0} = \frac{c_0}{c_2^2},
\end{equation*}
which leads to  error term \eqref{eq:cubicferror}.

\section{Synthetic data generation}\label{app:synthdata}
\subsection{Synthetic data generation}
Let $\mathcal U_{[a,b]}$ denote the uniform distribution in the interval $[a,b]$. 
In all experiments, the data set is generated as the uniformly sampled (period $0.5$) response to $4$ impulses. The time separation between the impulses, and the time from the last impulse to the end of the time horizon, have distribution $\mathcal U_{[2,5]}$. To obtain nonzero initial conditions, the impulse train is shifted so that time zero is situated at the midpoint between the first two impulses, and only the last $3$ impulses are included in the time series. 
The remaining parameters have distributions  according to Table~\ref{tab:datapars}.
\begin{table}[ht]
    \centering
    \caption{Distributions for impulse weights $d_k$ and elimination rates $b,b_1,b_2$ and standard deviations $\sigma$ of Gaussian additive noise. $\sigma_o$ is standard deviation of uniform noise representing outlying data points. Experiments are numbered according to the section they appear in.}
    \begin{tabular}{c|ccccc}
        Experiment & $d_k$ & $b_2$ or $b$ & $b_1-b_2$ & $\sigma$ & $\sigma_o$ \\
        \hline
        \ref{sec:firstorderexp} & $\mathcal U_{[0.1,1]}$ & $\mathcal U_{[0.4,1.4]}$ & & $0.01$ \\
        \ref{sec:secondorderexp} & $\mathcal U_{[0.4,4]}$ & $\mathcal U_{[0.4,1.4]}$ & $\mathcal U_{[0.3,1.3]}$ & See Table~\ref{tab:pointests} \\
        \ref{sec:basalexp} & $\mathcal U_{[2,7]}$ & $\mathcal U_{[0.4,1]}$ & $\mathcal U_{[4,5]}$ & $0.008$ \\
        \ref{sec:robustexp} & $\mathcal U_{[0.4,4]}$ & $\mathcal U_{[0.4,1.4]}$ & $\mathcal U_{[0.3,1.3]}$ & $0.006$ & $0.289$
    \end{tabular}
    \label{tab:datapars}
\end{table}

\end{document}